\documentclass{aa}  

\usepackage{txfonts}
\usepackage{graphicx}
\usepackage{amsmath}
\usepackage{amssymb}
\usepackage{natbib}
\bibpunct{(}{)}{;}{a}{}{,}
\usepackage{placeins}
\usepackage{float}
\usepackage{hyperref}
\usepackage{textcomp, gensymb}

\defcitealias{el-badry_what_2022}{EB+22}

\begin{document}

\title{The physical properties of post-mass-transfer binaries}

\author{R. Seeburger \inst{1,5} \and H-W. Rix \inst{1} \and K. El-Badry \inst{1,2} \and J. Müller-Horn \inst{1,5} \and A.~J.Dimoff \inst{4,1} \and J. Henneco  \inst{6,3}\and J.~I. Villase\~nor\inst{1}}

\institute{
    Max Planck Institute for Astronomy, Königstuhl 17, 69177 Heidelberg, Germany\\
        \email{seeburger@mpia.de}
\and
    California Institute of Technology, 1200 E California Blvd, Pasadena, CA 91125, United States
\and
    Heidelberg Institute for Theoretical Studies, Schloss-Wolfsbrunnenweg 35, 69118 Heidelberg, Germany 
\and 
    Institute of Applied Physics, Goethe University Frankfurt. Max-von-Laue-Str. 1, 60438 Frankfurt am Main, Germany
\and
    Universität Heidelberg, Department of Physics and Astronomy, Im Neuenheimer Feld 226, 69120 Heidelberg, Germany
\and
    School of Mathematics, Statistics and Physics, Newcastle University, Newcastle upon Tyne, NE1 7RU, United Kingdom
}

\date{December 2024}

\abstract
    {}
    {We present and analyse the detailed physical properties of six binary stellar systems, originally proposed as possible star-black hole binaries on the basis of radial velocities from Gaia's third data release, but soon recognised as likely post-mass-transfer binary systems with stripped companions.}
    {We used multi-epoch high-resolution FEROS spectra and spectral disentangling paired with stellar templates to derive effective temperatures, T$_\mathrm{ eff}$; stellar radii, $R_*$; and projected rotational velocities, v~$\sin{i}$ for both components in all systems along with the mass ratio, $q =  M_\mathrm{accretor} / M_\mathrm{donor}$ and the components' flux ratio as a function of wavelength.} 
    {Our analysis directly confirms that all systems are post-mass-transfer binaries with two luminous stars, i.e. no black hole companions. Each system contains an A-type accretor component that is rapidly rotating and a cooler very low-mass donor ($\sim 0.25 \mathrm M_\odot$) that is overluminous.
    Five of the systems show no trace of any emission lines, implying that there is no current mass transfer, consistent with our inferred radii, in all cases within the Roche volume. The data are generally consistent with stable case AB mass transfer with $\beta$ (the fraction of mass lost from the accretor) less than $0.7$.
    While the accretor components rotate rapidly, they rotate well below the critical rotation rate, $v_\mathrm{crit}$, even though there must have been enough mass transfer to spin them up to a significant fraction of $v_\mathrm{crit}$, according to theoretical models of angular momentum transfer. As neither magnetic braking nor tidal synchronisation should have been effective in spinning down the stars, our results suggest that either mass accretion does not increase the angular momentum of the accretors to their critical values or the systems never reached these values in the first place.}
    {}


   \maketitle

\section{Introduction}

\begin{table*}[!htb]
\caption{Results of the spectral analysis for the donor (don) and accretor (acc) for each system.}
\small
\centering
\begin{tabular}{llllllllll}
\hline
Gaia Source ID & Short ID & $T_\mathrm{eff, don}$ \tablefootmark{a} & $T_\mathrm{eff, acc}$ \tablefootmark{a}& $q$ \tablefootmark{a} & $T_\mathrm{eff, don}$ & $T_\mathrm{eff, acc}$ & $q_\mathrm{ TODCOR}$ & $v_\mathrm{rot, don}$ & $v_\mathrm{rot, acc}$ \\
 & & [kK] & [kK] & $ $ & [kK] & [kK] & $ $ & [km/s] & [km/s] \\
\hline
Gaia 2933630927108779776 & G-2933 & 4.4 $\pm$ 0.2 & 9.7 $\pm$ 0.3 & 8.1 $\pm$ 1.4& 4.75 $\pm$ 0.25 & 9.50 $\pm$ 0.25 & 5.7 $\pm$ 1.2 & 19 $\pm$ 1 & 176 $\pm$ 1 \\
Gaia 5694373091078326784 & G-5694 & 4.6 $\pm$ 0.2 & 9.9 $\pm$ 0.3 & 8.8 $\pm$ 1.7& 4.75 $\pm$ 0.25 & 9.25 $\pm$ 0.25 & 6.5 $\pm$ 1.1 & 21 $\pm$ 1 & 190 $\pm$ 1 \\
Gaia 6000420920026118656 & G-6000 & 4.5 $\pm$ 0.2 & 9.5 $\pm$ 0.3 & 11.6 $\pm$ 2.1& 4.75 $\pm$ 0.25 & 9.00 $\pm$ 0.25 & 5.1 $\pm$ 1.2 & 21 $\pm$ 1 & 156 $\pm$ 1 \\
Gaia 5243109471519822720 & G-5243 & 4.6 $\pm$ 0.2 & 9.4 $\pm$ 0.3 & 7.1 $\pm$ 1.2& 4.75 $\pm$ 0.25 & 8.75 $\pm$ 0.25 & 7.1 $\pm$ 2.4 & 20 $\pm$ 1 & 157 $\pm$ 1 \\
Gaia 2966694650501747328 & G-2966 & 4.7 $\pm$ 0.1 & 9.4 $\pm$ 0.2 & 8.7 $\pm$ 1.2& 5.50 $\pm$ 0.25 & 9.00 $\pm$ 0.25 & 5.4 $\pm$ 1.5 & 20 $\pm$ 1 & 164 $\pm$ 1 \\
Gaia 5536105058044762240 & G-5536 & 4.9 $\pm$ 0.2 & 8.4 $\pm$ 0.3 & 8.3 $\pm$ 1.7& 6.75 $\pm$ 0.25 & 7.25 $\pm$ 0.25 & 2.8 $\pm$ 0.7 & 12 $\pm$ 1 & 285 $\pm$ 1 \\
\hline
\end{tabular}
\label{tab:params}
\tablefoot{\tablefoottext{a}{Comparative values from \citetalias{el-badry_what_2022}.}
}
\end{table*}

\smallskip

Most massive stars and an appreciable fraction of lower-mass stars are born in binaries \citep{moe_mind_2017}. The initially more massive star evolves first and expands, potentially leading to mass transfer towards its companion. Such mass transfer can fundamentally alter both components and the whole configuration of the system, as the donor loses most of its envelope and becomes a low-mass stripped star, while the accretor gains mass and is spun up by the angular momentum of the accreted material \citep{packet_spinup_1981a}. The outcome of this interaction depends critically on the initial orbital period, component masses, and the efficiency with which mass and angular momentum are accreted \citep{packet_spinup_1981a,soberman_stability_1997a}.

For a brief period, the low-mass stripped star can become very luminous, outshining its far more massive (accretor) companion star in the binary. During this phase, the high inferred mass ratios (and thus, large implied secondary mass) and the lack of obvious evidence of a secondary star due to rotationally broadened spectral lines, can make these systems appear as black hole binary `impostor' systems \citep{shenar_hidden_2020a, bodensteiner_hr_2020a, el-badry_ngc_2022b,el-badry_ngc_2022c,el-badry_unicorns_2022}.
Immediately after the Gaia data release 3 (DR3)\citep{gaiacollaboration_gaia_2023c}, \cite{el-badry_what_2022} (hereafter \citetalias{el-badry_what_2022}) analysed a sample of Gaia DR3 binaries initially flagged as possible black hole hosts based on their high mass functions. By modelling their spectral energy distributions (SEDs) and light curves, \citetalias{el-badry_what_2022} showed that these systems were actually systems containing a stripped donor star and a main-sequence accretor.

These systems containing a relatively hot ($\sim9000$~K) accretor and a cooler ($\sim5000$~K), less massive ($\lesssim 0.5$ M$_\odot$) but luminous ($\gtrsim 10$ L$_\odot$) donor, represent an important stage in binary stellar evolution. When seen edge-on, they may appear as Algol-type variables, where paradoxically the less massive component is initially the more luminous of the two \citep{kopal_classification_1955}. This is due to previous mass transfer in the system, which reversed the mass ratio \citep{kippenhahn_entwicklung_1967b}. While the basic physical picture of how such systems form through mass transfer is understood, many aspects remain poorly constrained empirically. In particular, how conservative mass transfer is (i.e. how much mass is retained in the binary versus lost to the interstellar medium) \citep[e.g.][]{demink_efficiency_2007a, claeys_binary_2011a, deschamps_criticallyrotating_2013}, mass transfer stability \citep[e.g.][]{Hjellming1989a, Hjellming1989b, Deloye2010, Ge2010, Ge2010b, Ge2015, Ge2023, Ge2024, Woods2011, Pavlovskii2015, temmink_coping_2023a, Ivanova2024}, the mechanisms governing stellar spin during and after mass transfer \citep[e.g.][]{sun_stellar_2024}, and the detailed physical properties of stripped donor stars \citep[e.g.][]{gotberg_stellar_2023a} constitute some of the biggest uncertainties in binary evolution while tremendously affecting the outcome of binary evolution models \citep[e.g.][]{marchant_evolution_2024a, pols_mass_2007}.

Here, we present a detailed spectroscopic follow-up study of six systems using multi-epoch high-resolution spectra taken with the
Fiber-fed Extended Range Optical Spectrograph \citep[FEROS,][]{kaufer_commissioning_1999} instrument. Out of the sample of 14 objects identified by \citetalias{el-badry_what_2022}, six are at declinations inaccessible to FEROS, while of the remaining eight, six were at right ascensions observable by FEROS during the observing period applied for, P112. Thus, these six are the targets analysed in this work, and they are listed in Table \ref{tab:params}. For brevity, we refer to them in the text by their first four numbers only (i.e. Gaia 2933630927108779776 is G-2933).

By applying the spectral disentangling method (section \ref{subsec:disentangling}) and employing stellar spectral templates, we derived fundamental parameters (effective temperature, $T_\mathrm{ eff}$; stellar radius, $R_*$; and projected rotational velocity, $v\sin{i}$) for both components, along with mass ratios, $q$; and wavelength-dependent component flux ratios. This allowed us to constrain the current physical state of these systems and probe their mass transfer history. We build on the work of \citetalias{el-badry_what_2022}, to provide not only spectroscopic stellar parameters but also clarify whether mass transfer is still ongoing, how conservative the mass transfer must have been, and how much rotation the accretor exhibits. To test these expectations and characterise the present-day properties of these systems, we obtained new, multi-epoch, high-resolution spectroscopy, which we describe in the following section.

\section{Observations and data reduction}
\label{sec:obs}

We acquired multi-epoch spectroscopic data \footnote{ID: 112.264Z.001, PI: Seeburger} using the FEROS instrument at the MPG/ESO 2.2m telescope at La Silla, Chile. It is located at 29°15'28" S, 70°44'12" W, at an altitude of 2375 m. The telescope uses a Ritchey-Chrétien reflector, with the FEROS instrument at Cassegrain focus, covering a wavelength range of $\sim3500$ \AA~ to $\sim9200$ \AA~ at a resolution of $\sim48,000$.

Observation details are consolidated in table \ref{tab:obs} in the Appendix. We observed each target about ten times with a one or two nightly cadence, depending on the system's reported periods in \citetalias{el-badry_what_2022}. As all targets have orbital periods of $\sim$10 to 20 days, this cadence provides good coverage of the RV dynamical range.
We used exposure times of 20-30 min to reach a S/N of about 40-150, depending on the $G$-band magnitude as reported by Gaia (see Table \ref{tab:orbital} in the Appendix).

The FEROS instrument is well-suited to this kind of analysis, as it provides good resolution over a wavelength range that includes the most important Balmer lines, relevant for our analysis of the hotter accretor, as well as many metal lines. We reduced data from FEROS using the \verb|ceres| pipeline \citep{brahm_ceres_2017}, which returns spectra in the wavelength range $\sim3900$ \AA~ to $\sim6770$ \AA~.

After acquisition and reduction of the data, we normalised the observed spectra with a running median filter. \verb|ceres| provides normalised spectra, but we used the unnormalised spectra and performed our own normalisation to ensure the data and model spectra were normalised by the same method. Additionally, we created a set of template spectra by normalising and resampling \cite{coelho_new_2014} model spectra in the relevant temperature, metallicity, and $\log{g}$ ranges onto the same wavelength grid as the FEROS data. The template spectra were computed assuming local thermal equilibrium and spaced in intervals of 250 K ($T_\mathrm{ eff}$), and 0.5 ($\log{g}$). Here, we consider temperatures from 4000 K to 11000 K, and surface gravities from two to four. We assumed solar metallicity and applied rotational broadening with the \verb|RotBroadInt| package by \cite{carvalho_simple_2023} from 0 km/s to 300 km/s.

Using these data, our next goal was to derive the stellar and orbital parameters of both components. In the following, we outline the sequence of analyses performed, each building on the results of the previous step.

\section{Data modelling and results}
\label{sec:method}

We present the methodology and observational results for our sample in this section. 
This is followed with a comparison of our results with \citetalias{el-badry_what_2022} and a discussion of the significance for the physical state and history of these systems in the subsequent Section~\ref{sec:discussion}.

\subsection{Spectral disentangling and determining RVs}
\label{subsec:disentangling}

\begin{figure*}[p]
    \centering
    \includegraphics[width=0.49\linewidth]{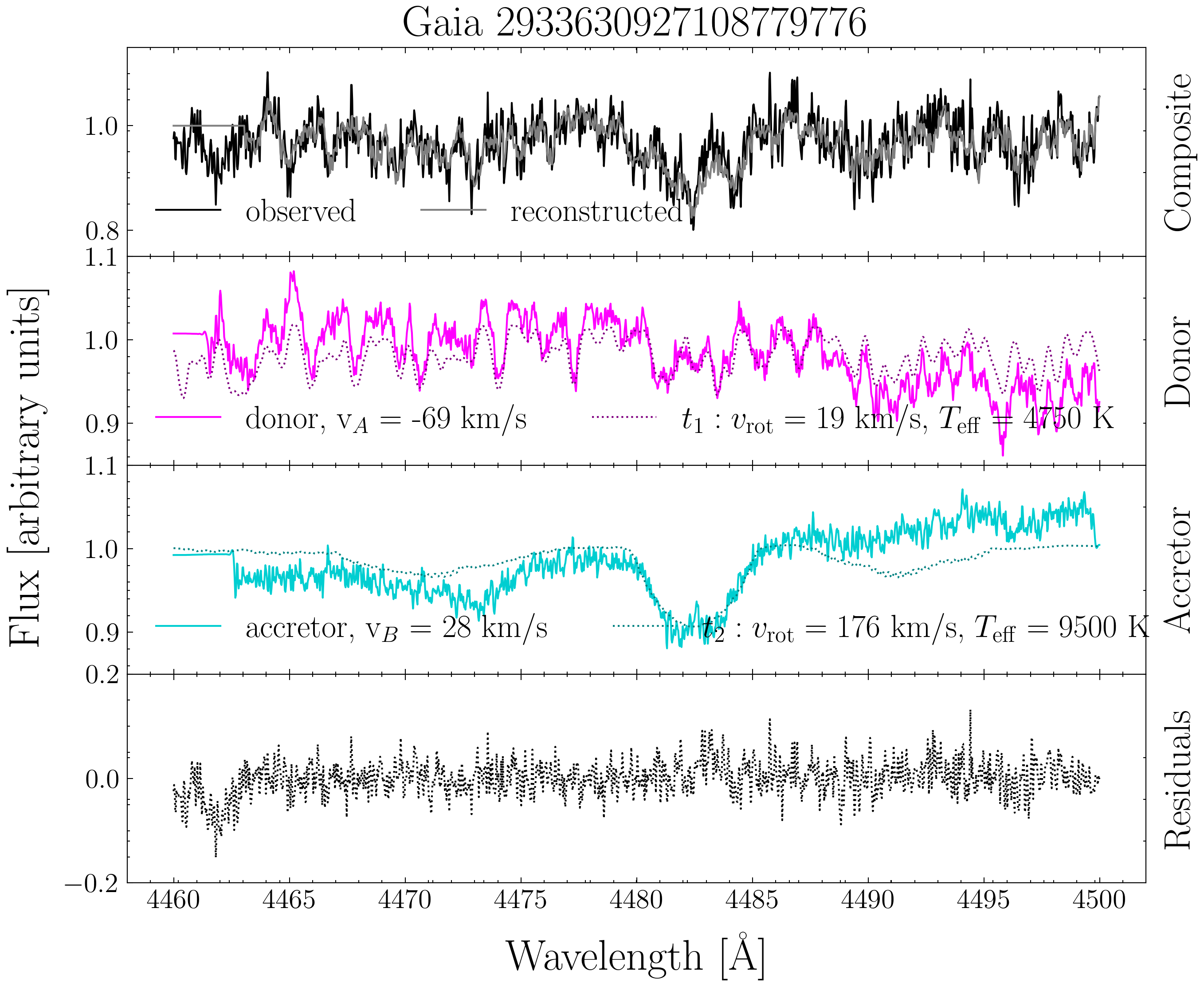}
    \includegraphics[width=0.49\linewidth]{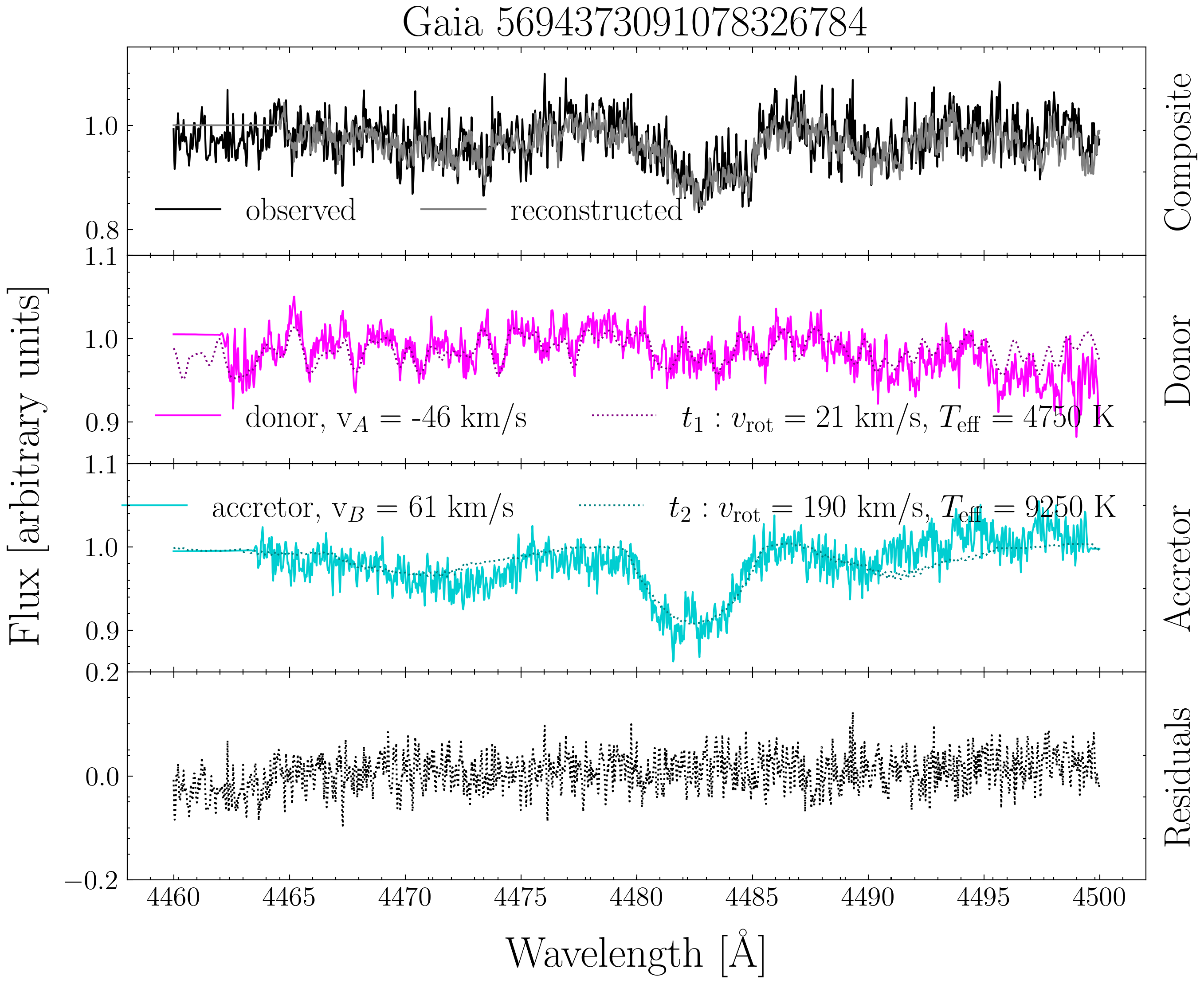}
    \includegraphics[width=0.49\linewidth]{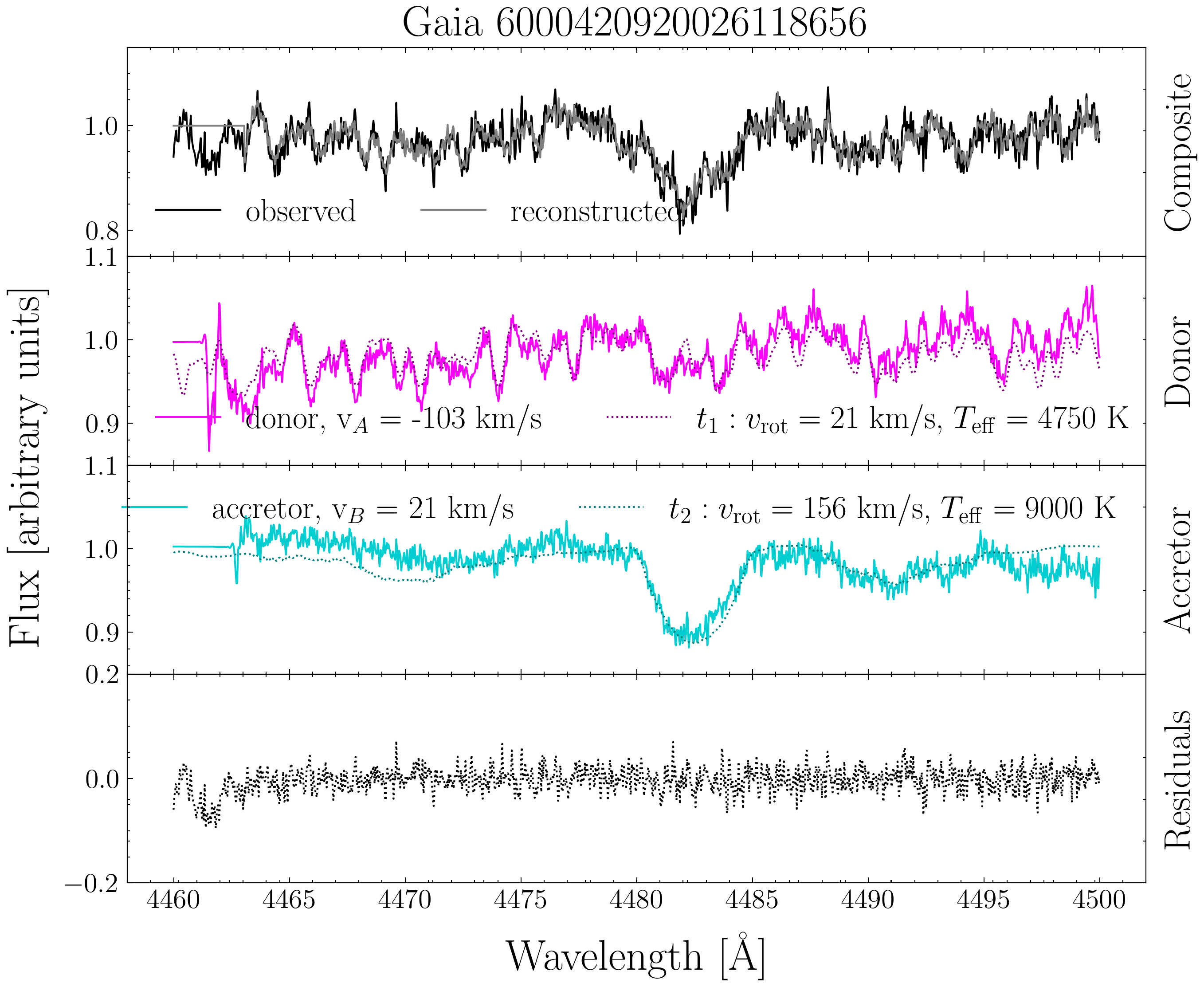}
    \includegraphics[width=0.49\linewidth]{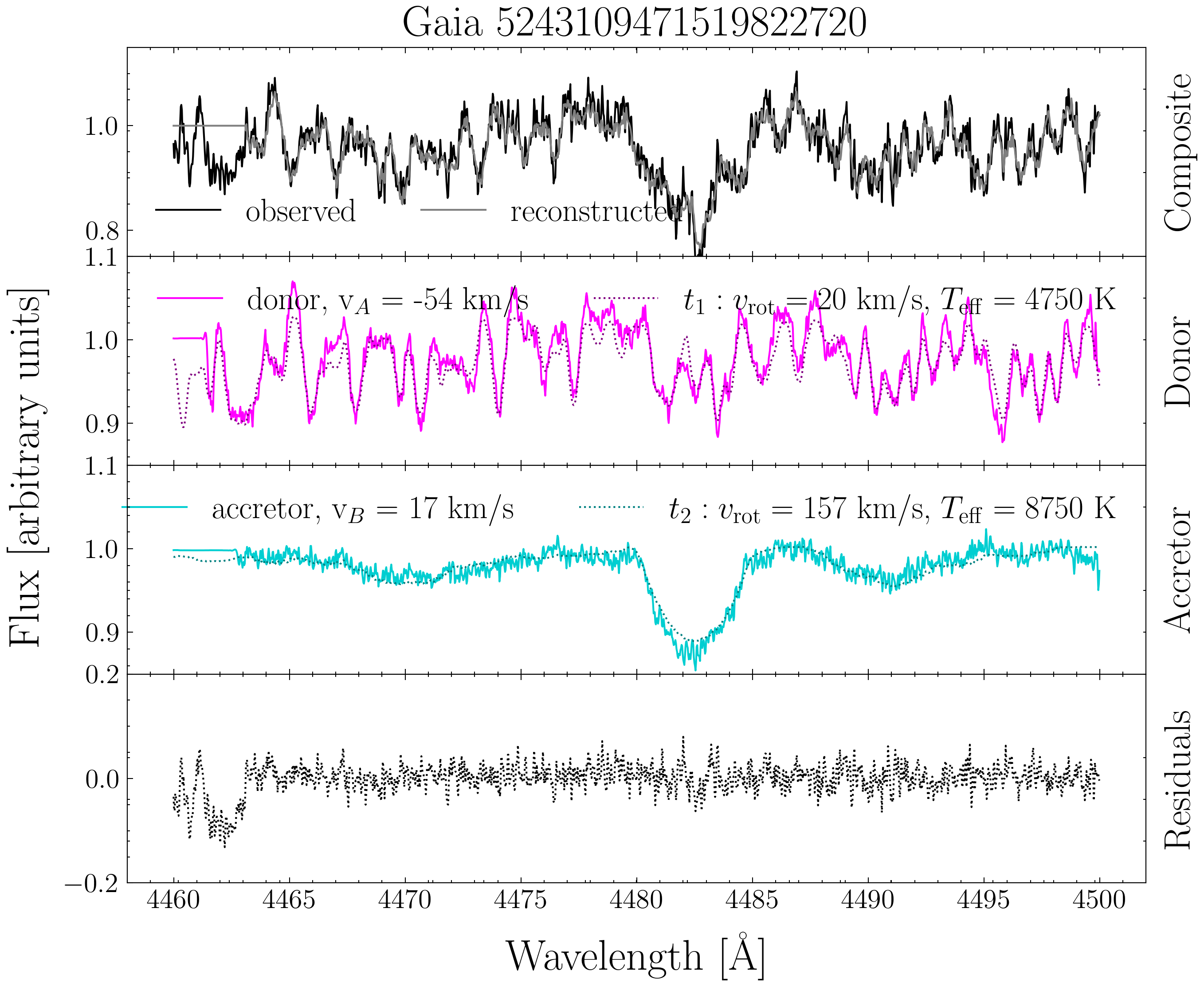}
    \includegraphics[width=0.49\linewidth]{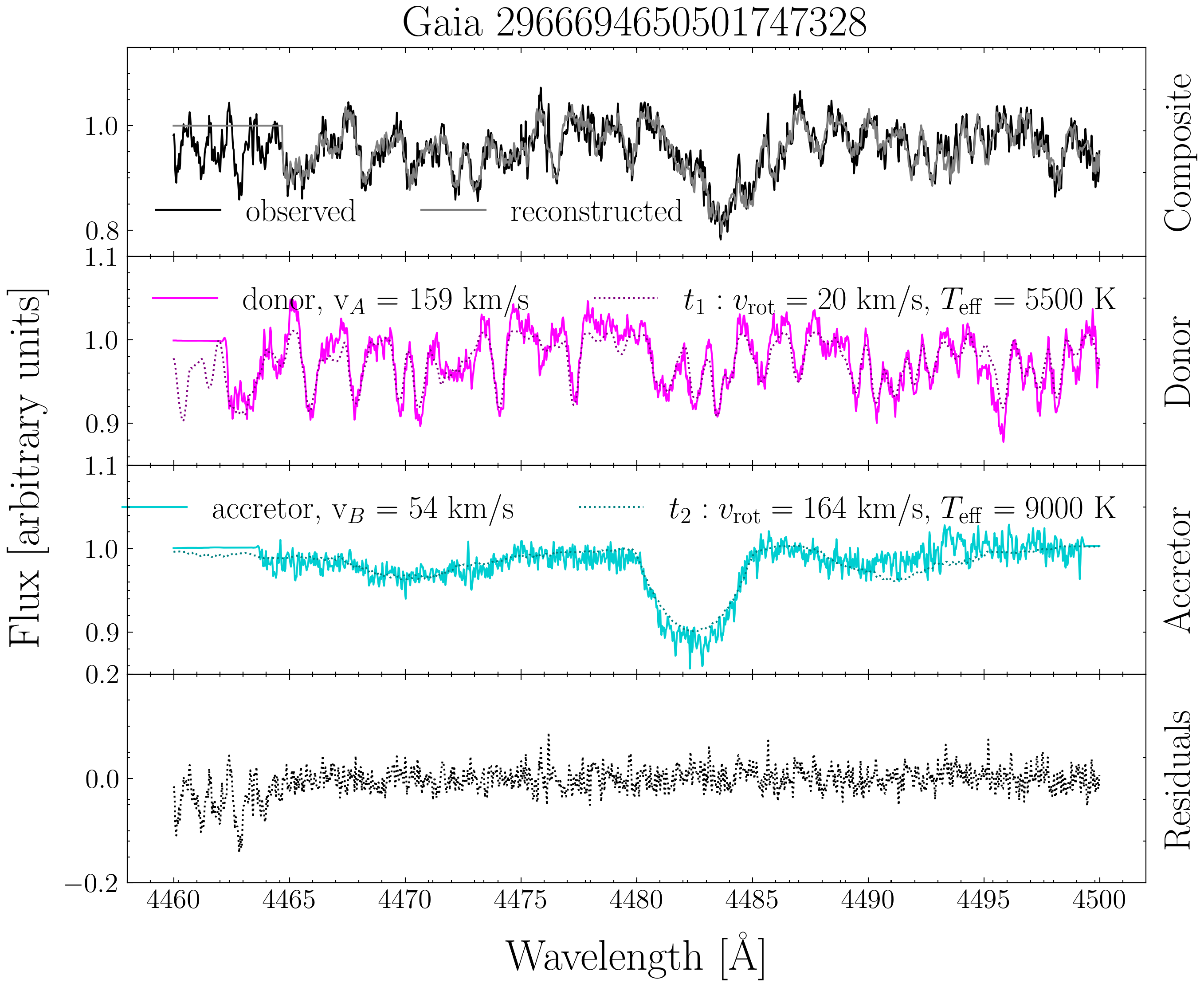}
    \includegraphics[width=0.49\linewidth]{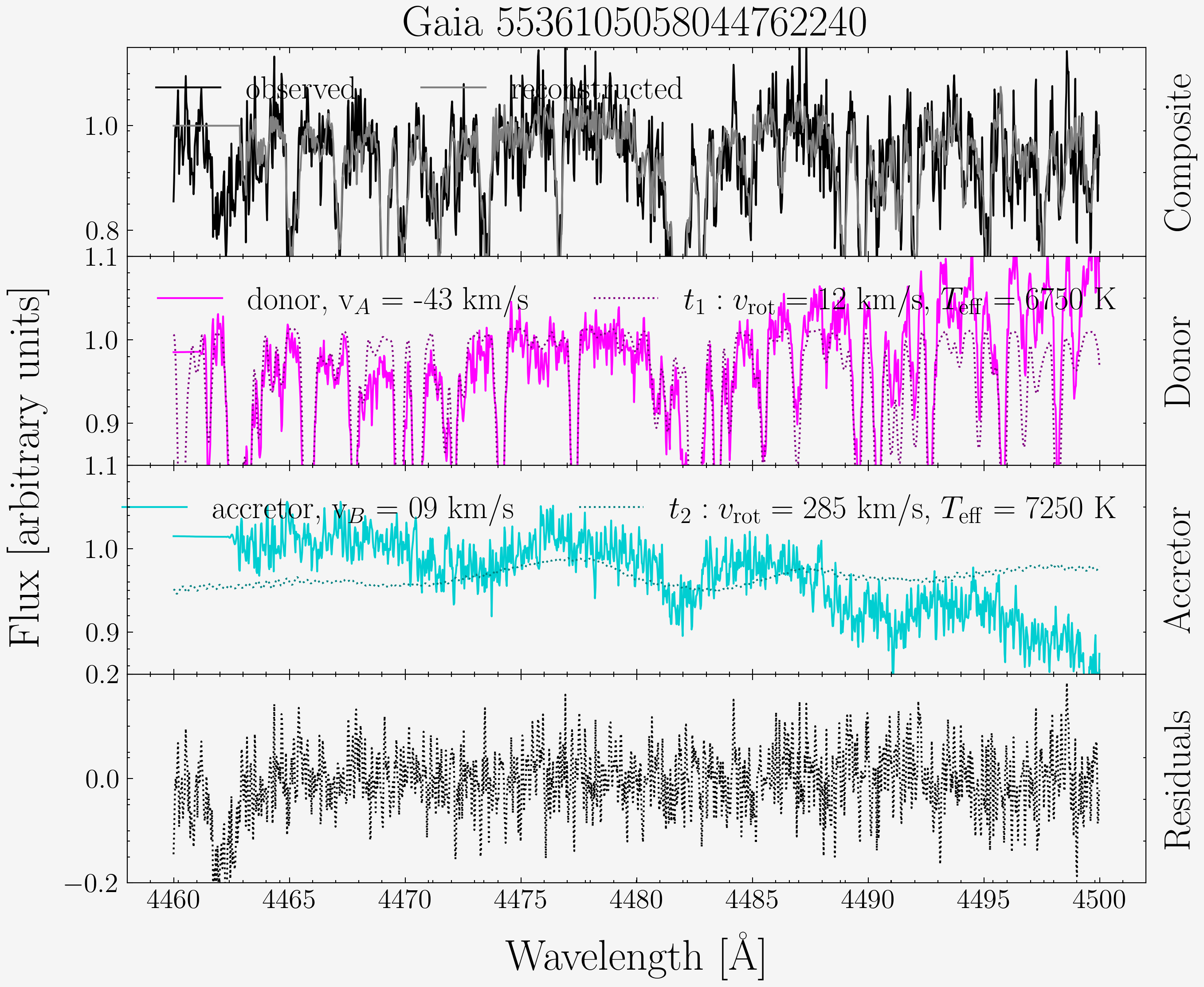}
    \caption{Observed spectra for one epoch (black line, top row), the two disentangled components (magenta and cyan lines, middle row, shown in the rest frame), their sum (grey line, top row, computed by shifting and co-adding), and the residual (bottom row) for the six targets of this study, centered around the \ion{Mg}{ii} line as seen clearly in the accretor spectrum. We also show the best-fit template spectra for each disentangled component in the middle rows (dotted lines). In this and following figures, plots showing results for target G-5536 have a light grey background to differentiate them from the other targets, due to the difficulties with the analysis encountered for this object.}
    \label{fig:initspectra}
\end{figure*}

\begin{figure*}[!tb]
    \centering
    \includegraphics[width=1\linewidth]{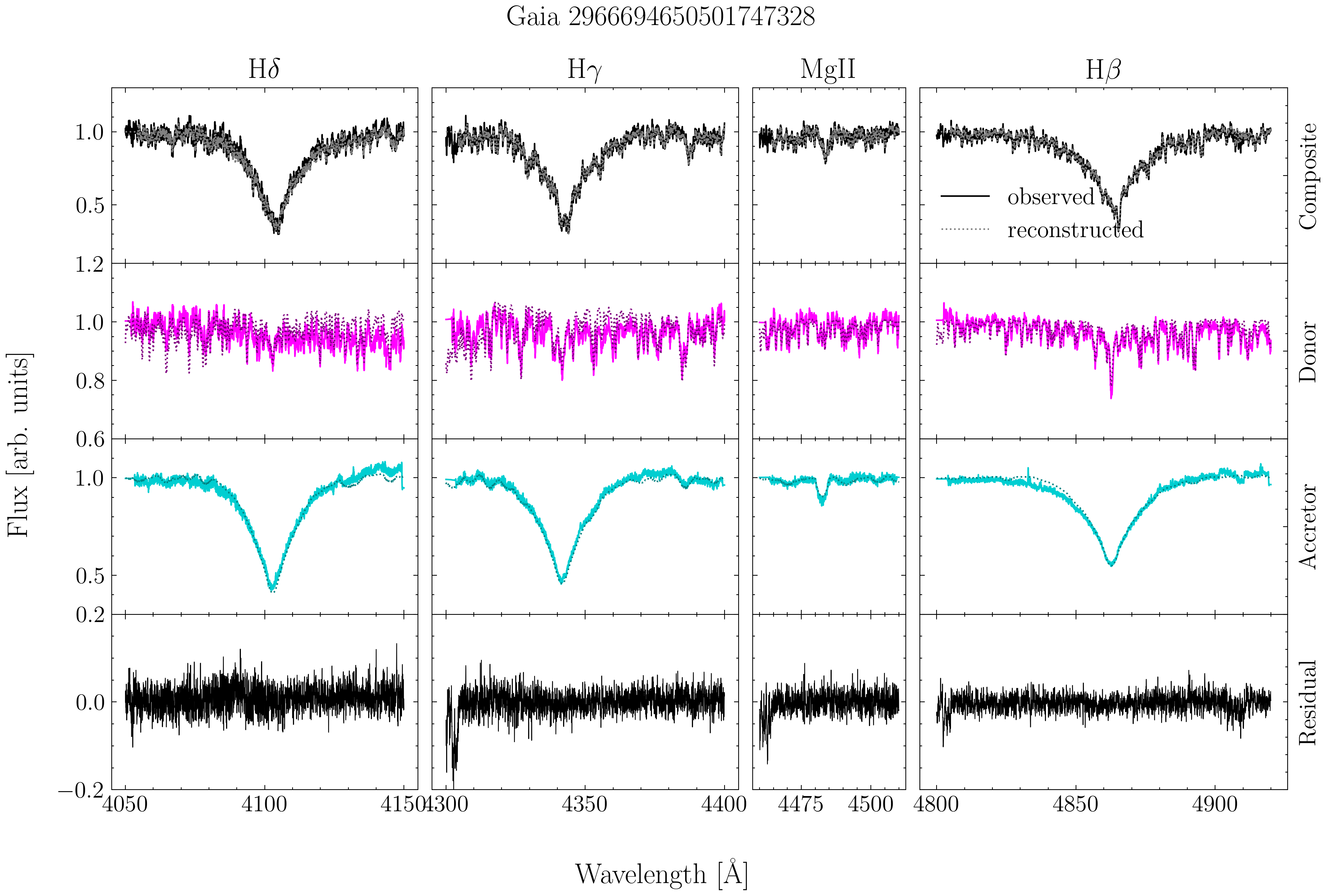}
    \caption{Observed spectra for one epoch (black line, top row), the two disentangled components (magenta and cyan lines, middle rows), their sum (grey line, top row), and the residual (bottom row) for a number of different wavelength ranges. We also show the best-fit template spectra for each disentangled component in the middle rows (dotted lines). The windows here show the Balmer series and \ion{Mg}{ii} in the accretor, which is hot and has few lines.
    Here, we only show the spectra for one object (G-2966), the rest can be found in Figures \ref{fig:regions1} and \ref{fig:regions2} in the Appendix.}
    \label{fig:regions}
\end{figure*}

As a first step, we disentangled the composite spectra into the individual component spectra. This provided the foundation for all subsequent analyses by yielding both the radial velocities (RVs) and the individual stellar spectra.

In this work we make extensive use of the spectral disentangling pipeline described in \cite{seeburger_autonomous_2024}, based on \cite{simon_disentangling_1994}. Spectral disentangling is a data-driven method that seeks to decompose multi-epoch observational spectra of a multiple system into the individual component spectra. By assuming that the component spectra are time invariant in their shape and only change by being red- and blue-shifted, one can construct a matrix that, when multiplied by the two component spectra, reconstructs the observations at each epoch. The values of the matrix elements are determined by the RVs of each component at each epoch. This can be simplified by first determining the primary (donor) velocities and the mass ratio of the system and then computing the secondary (accretor) velocities from these (see Section 2.5 in \cite{seeburger_autonomous_2024} for details). 

\cite{seeburger_autonomous_2024} combined this linear algebra problem with a non-linear optimisation step. At each step of the optimiser, a vector of primary RVs and a mass ratio is proposed, the spectra are disentangled, and the residuals between the multi-epoch observations and reconstruction are computed. This residual is taken to be a metric of fit and sought to be minimised by the optimiser, in theory returning the parameter set for which the algorithm could most accurately decompose the spectra, as well as the best-fit component spectra.

In this work, we computed initial guesses for the primary RVs by cross-correlating the observed spectra with our templates. Then, we iteratively optimised the primary velocities and mass ratio by proposing a set of candidate parameters, disentangling and computing the residual for that set, seeking to minimise the residual. We found the region around H$\beta$ to be optimal for this, as both spectra contribute similarly to the total flux, and there are sufficient features in both components (abundant metal lines in the donor, H$\beta$ in the accretor) to perform the disentangling. This is particularly important for the accretor, which is hot and has only a few lines. 

The basic result of this analysis are the disentangled spectra for all six systems, with velocities and mass ratio determined via optimisation. These disentangled spectra are shown in Figure~\ref{fig:initspectra}, for the \ion{Mg}{ii} wavelength region, where, in all six systems, both components show prominent features. For each of the six panels, the top displays the total spectrum at one epoch, the next two the two disentangled components (donor and accretor), and the bottom panel shows the residuals. Figure~\ref{fig:initspectra} demonstrates that all systems yield well-defined and physically plausible disentangled solutions that leave only small systematic residuals. This immediately confirms that all these systems are indeed SB2 (double-lined spectroscopic binaries, as opposed to SB1s, single lined spectroscopic binaries) binary systems with two luminous components, as surmised by \citetalias{el-badry_what_2022}.

All panels of Figure~\ref{fig:initspectra} show that the primary spectrum (magenta, second panel) exhibits an abundance of narrow metal absorption lines, characteristic of cool stars without fast rotation. The secondary spectrum (cyan, third panel) shows far fewer and much broader lines, indicative of a hotter and rapidly rotating star. Qualitatively, this is aligned with the picture where the cool primary star is a low $\log{g}$ donor, while the hotter secondary star is the spun-up accretor. The nature of the hotter disentangled component is illustrated further in Figure~\ref{fig:regions} for one of the objects (G-2966): the Figure zooms in on three Balmer line sub-regions (H$\beta$ to H$\delta$ - H$\alpha$ is not shown due to stationary, telluric lines near the stellar line affecting the disentangling), in addition to the \ion{Mg}{ii} region shown already in Figure~\ref{fig:initspectra}. The Balmer lines are the dominant spectral features of the hotter component, indicative of A-type stars with $T_\mathrm{eff} \approx 9000$ K. Figure~\ref{fig:regions} also illustrates that our sample star (G-2966) shows no wings on either side of the H$\beta$ line, indicating little or no H$\beta$ emission. This affirms expectations for a system currently not undergoing mass transfer. Analogous figures for the other five targets are shown in Figures \ref{fig:regions1} and \ref{fig:regions2} in the Appendix.

\subsection{Stellar parameter determination}
\label{subsec:stellar_paran}

Having obtained disentangled spectra for both components, we next determined their effective temperatures and projected rotational velocities. These parameters provide the basis for assessing the physical state of each star and for subsequent comparisons with evolutionary models.

First, we focused on the disentangling result around the red end of the spectrum (4800 - 5700 \AA), where the cooler donor dominates the flux, allowing us to fit a template to the disentangled donor spectrum. We did so by considering the template spectra described above, at all available effective temperatures and surface gravities, but only a selection of rotational velocities, as these were determined in a subsequent step. We scaled each template by a set of different light ratios between 0.01 and 100, and performed a least squares fit with the disentangled donor spectrum. We repeated the same procedure in the blue (3950 - 4900 \AA) for the hotter accretor. This provided the effective temperatures and surface gravities of each component from the corresponding best-fit template. Additionally, by considering the disentangled spectra around the \ion{Mg}{ii} line at 4481 \AA, we determined the projected rotational velocity of the stars, crucial to understanding the spin-up and/or spin-down that has occurred during mass transfer. We achieved this by selecting the best-fit template in terms of $T_\mathrm{eff}$ and $\log{g}$ from the previous step (where we considered only a handful of potential rotational velocities), and considered all rotationally broadened instances thereof, subsequently determining the best-fit rotational velocity via a least-squares fit of the disentangled spectra and broadened templates. We preferred \ion{Mg}{ii} over one of the (deeper) Balmer lines, as the shape of the Balmer lines is predominantly set by the stellar temperature and surface gravity, and rotation has only a comparatively small effect.

Table \ref{tab:params} lists the best template temperatures and rotation velocities for both components. Further, these are also shown on the relevant panels of Figures \ref{fig:initspectra} and \ref{fig:regions}, indicating the stellar parameters of the template (dotted lines) displayed in that panel. All donors show significant but moderate rotational velocities consistent with tidal synchronisation. The accretors rotate more rapidly, but in all cases well below critical rotation, $v_\mathrm{crit}$, the surface velocity of the star where its outward centrifugal force is equal to the inward force exerted by gravity, also known as break-up velocity. This can be computed as:

\begin{equation}
    v_\mathrm{crit} = \sqrt{\frac{G M_*}{R_*}},
\end{equation}

where $M_*$ and $R_*$ are the mass and radius of the star in question, and $G$ is Newton's constant. For a detailed discussion, see Section \ref{sec:discussion}. We then used these best templates for subsequent analysis.

\subsection{Applying TODCOR for an independent mass ratio estimation}
\label{subsec:TODCOR}

To independently verify and refine the mass ratios inferred from the disentangling procedure, we also applied the TwO-Dimensional CORrelation \citep[TODCOR, ][]{zucker_study_1994} method. The mass ratio is an essential parameter for the mass transfer history analysis in Section \ref{subsec:MT_history}.

We used the best-fit templates for each component with TODCOR to determine the velocities of the two components for each epoch. The primary and secondary velocity ($v_A$ and $v_B$, respectively) at each epoch were chosen such that they maximised the value of the 2D cross-correlation of the light-ratio-scaled templates with the observed spectrum at that epoch. The primary and secondary velocity are related by 
\begin{equation}
\label{eq:RVB}
    v_B =  v_\mathrm{COM} + \frac{v_\mathrm{COM} - v_A}{q},
\end{equation}
where $v_\mathrm{COM}$ is the centre-of-mass velocity. This shows that $v_B$ should be a linear function of $v_A$ with a slope and intercept of 
\begin{align}
    \label{eq:slopeintercept}
    \text{slope} = -\frac{1}{q}
    && 
    \text{and} 
    &&
    \text{intercept} = v_\mathrm{COM} + \frac{v_\mathrm{COM}}{q}.
\end{align}
We fit this straight line using scipy's \verb|curve_fit| \citep{virtanen_scipy_2020a}, a non-linear least squares algorithm to fit a function to data, and derived the desired quantities. \verb|curve_fit| also returns the covariance matrix, which allowed us to determine the error on the parameters. Thus, TODCOR provided an independent measurement of the mass ratio and centre-of-mass velocity from the individual epoch velocities. 

\begin{figure*}[!htb]
    \centering
    \includegraphics[width=0.49\linewidth]{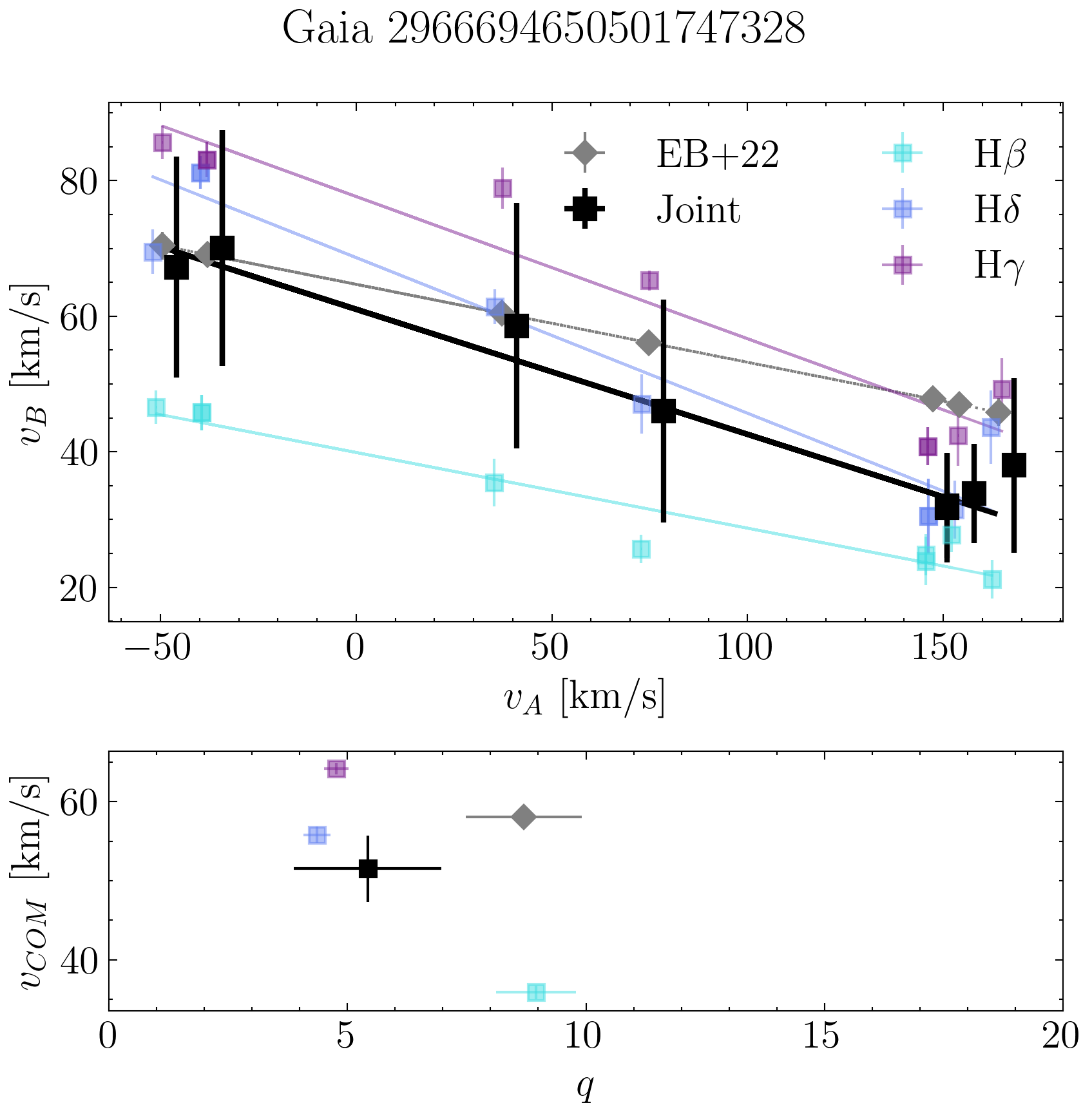}
    \includegraphics[width=0.49\linewidth]{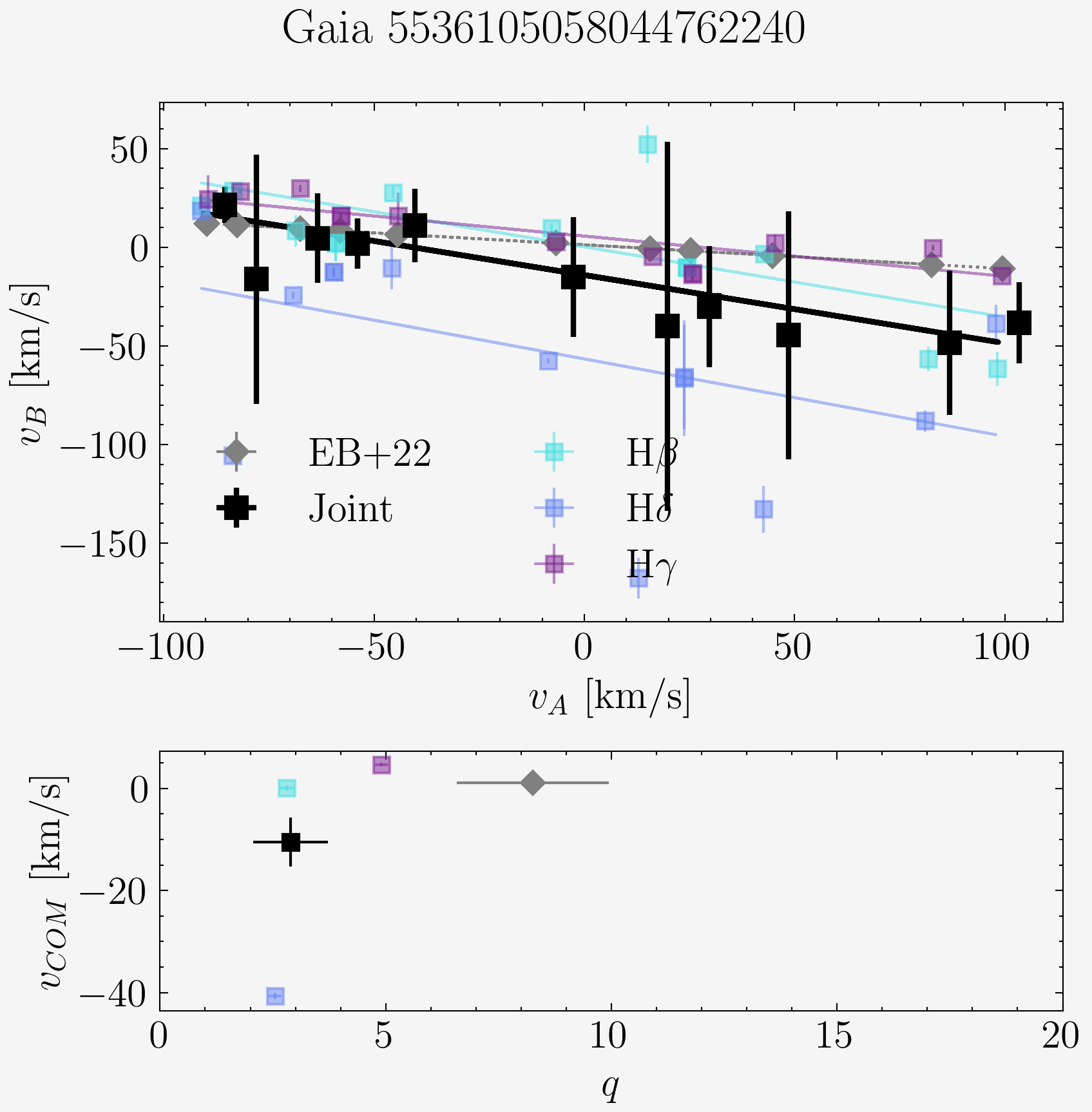}
    
    \caption{
    Results of the TODCOR algorithm applied to two objects. Top panels: RV of the donor and accretor shown on the x- and y-axes, respectively. Markers show the RVs of both components determined for each epoch with different methods and in various wavelength windows, with each marker displaying a different epoch. The lines indicate the linear best-fit to the RVs, as the relationship between donor and accretor RVs is linear. Grey diamonds show the donor RVs from \texttt{ceres} \citep{brahm_ceres_2017} on the x-axis, with corresponding accretor RVs on the y-axis. The accretor RVs were computed using the \texttt{ceres} donor RVs, the mass ratio from \citetalias{el-badry_what_2022}, centre-of-mass velocity from Gaia, and equation \ref{eq:RVB}. Coloured squares and lines show the results from TODCOR for a number of wavelength windows centered on different lines. Black squares show the unweighted mean for each epoch velocity over the different wavelength windows, offset to higher $v_A$ by a small amount for clarity. The fit labelled `Joint' is then performed to these mean values. From the slope and intercept of the best-fit line, the mass ratio and centre-of-mass velocity can be computed, see equation \ref{eq:slopeintercept}. Bottom panels: Plot of the mass ratio vs the centre-of-mass velocity as determined from each set of data in the top panel, coloured accordingly. 
    The left panel shows target G-2966, while on the right we see object G-5536, with a grey background to highlight the difficulty of the analysis for this object. For G-5336, we see that while there is a big discrepancy between the mass ratio from TODCOR and the one found by \citetalias{el-badry_what_2022}, the quality of the linear fit with TODCOR is not good, calling this result into question.}
    \label{fig:TODCOR}
\end{figure*}

We illustrate this for two of the objects (G-2966 and G-5536) in the top left and right panels of Figure~\ref{fig:TODCOR}, respectively. The panels show the component velocities, estimated from a range of different wavelength windows. The velocity estimates for the cool, narrow-lined donor component (along the x-axis of Figure~\ref{fig:TODCOR}) are consistent across wavelength windows with very small uncertainties. For the accretor, however, the velocity estimates vary substantially and are inconsistent across the different wavelength windows, and their formal uncertainties are considerably larger than those of the donor velocities.
Indeed, these (presumably also systematic) uncertainties made the determination of the mass ratio -- from the slope of the best-fit line -- difficult for all objects. The different values of the mass ratio and centre-of-mass velocity are shown for the corresponding lines in the bottom panels, illustrating the discrepancies. We discuss further in Section \ref{sec:discussion}. Our best estimates of the dynamical mass ratios, $q$, are listed in Table~\ref{tab:params}.

\subsection{Finding the light ratio as a function of wavelength}
\label{subsec:flux_ratio}

The relative flux contribution of each component carries additional information about the system’s configuration. We therefore determined the wavelength-dependent light ratio, linking the spectroscopic analysis with the photometric SED modelling that followed.

To do so, we considered the disentangled spectra across the \verb|ceres| spectral range, split into 20 wavelength windows. For each window, we wished to ascertain the optimal light ratio between the (previously determined as described in Section \ref{subsec:stellar_paran}) best-fit donor and accretor templates. To this end, we scaled the donor and accretor templates by a factor 

\begin{align}
    l_\mathrm{ don} = \frac{1}{1+a} =  \frac{f_\mathrm{donor}}{f_\mathrm{total}}
    &&
    \mathrm{and} 
    &&
    l_\mathrm{ acc} = \frac{a}{1+a} = \frac{f_\mathrm{accretor}}{f_\mathrm{total}} 
\end{align}

respectively, computing the best fit via least-squares minimisation between the disentangled spectra and the respective scaled templates. This allowed us to determine the optimal light ratio 

\begin{equation}
    a = \frac{f_\mathrm{accretor}}{f_\mathrm{donor}},
\end{equation}

and, trivially, the relative light contribution of the donor $l_\mathrm{ don}$ for the window in question. Using this method, we got a piecewise approximation of the light ratio as a function of wavelength, which subsequently provided an additional constraint in our SED fits.

\begin{figure*}[p]
    \centering
    \includegraphics[width=\linewidth]{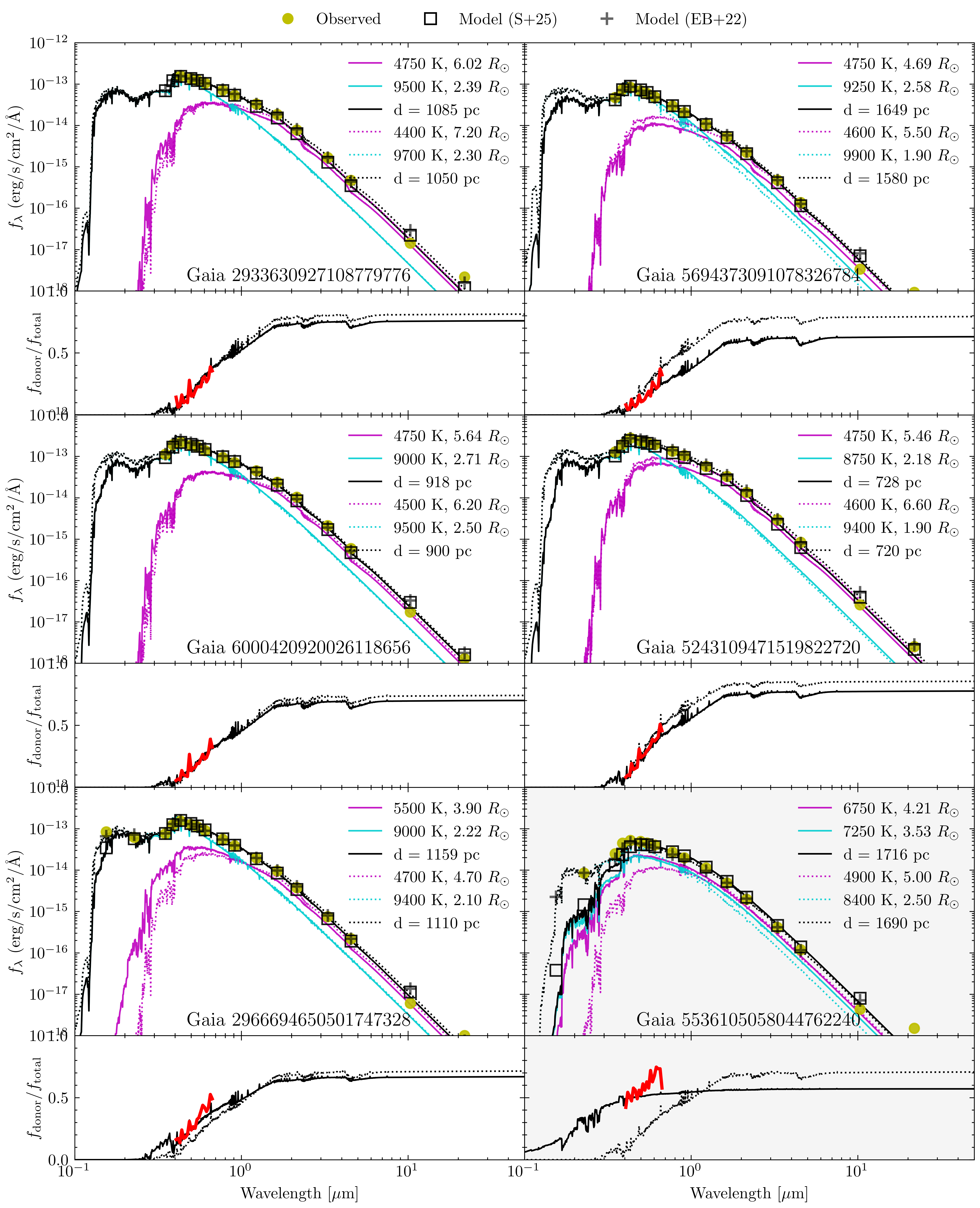}
    \caption{Spectral energy distribution fits and light ratio as a function of wavelength for each target. In the top plot of each set, we see  the donor's (magenta line) and accretor's (cyan line) model SED, as well as sum of the two component models (black line). Solid lines use stellar parameters from this work, and dotted lines the parameters derived in \citetalias{el-badry_what_2022}. We also overplot the observed photometry (lime dots), as well as the mock photometry from this work (black squares) and \citetalias{el-badry_what_2022} (grey crosses). Subsection \ref{subsec:sedfit} describes how each of these were obtained/computed. The bottom panel shows the contribution of the donor to the total flux as a function of wavelength. Again, solid indicates this work, dotted \citetalias{el-badry_what_2022}. The red line shows the spectroscopic light ratio derived in this work, providing an additional constraint in the SED fitting process. Target G-5536 is shown with a grey background to highlight the difficutlies with its analysis.}
    \label{fig:SEDs}
\end{figure*}

The donor spectrum's contribution to the total flux, $l_\mathrm{ don}(\lambda) = f_\mathrm{donor}/f_\mathrm{total}$ is shown for all objects in the bottom sub-panels of Figure~\ref{fig:SEDs} (red line). This Figure shows that in all cases the donor flux contribution increases strongly towards longer wavelengths, as expected from the much cooler temperature of the donor. This also explains why the disentangled donor spectra at short wavelengths in Figure~\ref{fig:regions} appear so noisy: The total flux at the bluest wavelengths is dominated by the accretor.

\subsection{SED fitting}
\label{subsec:sedfit}

To connect the spectroscopic and photometric constraints and to verify the consistency of the derived parameters, we modelled the SEDs of the systems using multi-band photometry. SED fits were initially used by \citetalias{el-badry_what_2022} to identify the systems studied here as potential SB2s, without priors on the stellar parameters from spectroscopy. We followed a similar method as the original authors. 

We acquired observed photometry for each target from the Wide-field Infrared Survey Explorer \citep[WISE, ][]{wright_widefield_2010}, SkyMapper \citep{keller_skymapper_2007}, American Association of Variable Star Observers (AAVSO) All-Sky Photometric Survey \citep[APASS, ][]{henden_vizier_2016}, the Galaxy Evolution Explorer \citep[GALEX, ][]{martin_galaxy_2005} and the Two Micron All Sky Survey \citep[2MASS, ][]{skrutskie_two_2006}, covering a wide wavelength range. For each component of the system, we simulated synthetic spectra with \verb|pystellibs| \citep{fouesneau_pystellibs_2016}, and added them to obtain a composite model SED. Dust attenuation was then performed using the \verb|dustmaps| package \citep{green_dustmaps_2018} with the 3D dustmap from \cite{edenhofer_parsecscale_2024}. We created mock photometry using \verb|pyphot| \citep{fouesneau_pyphot_2024}. This package contains a library of common photometric filters, including their throughput as a function of wavelength. Then, for a given flux through a filter, the total number of photons, and thus energy flux, can be computed via integration. More detail is provided in the \verb|pyphot| documentation \footnote{https://mfouesneau.github.io/pyphot/photometry.html}.

Then, using the Markov chain Monte Carlo sampling package \verb|emcee| \citep{foreman-mackey_emcee_2013}, we determined the best set of parameters for each component of the SED to reconstruct the observed photometry. Here, we held the effective temperatures fixed to the values acquired from the spectroscopy and set the surface gravities to the values inferred by \citetalias{el-badry_what_2022}. We placed a Gaussian prior on the parallax based on the Gaia DR3 parallax and its error, and a flat prior on the radii for both components. Thus, we performed a three-parameter fit: the parallax, radius of the donor and radius of the accretor. 

We then computed the flux ratio of the two simulated components across the wavelength range covered by the FEROS observations.  Comparing it to the spectroscopically determined flux ratio leads to an additional term in our full likelihood function:

\begin{equation}
    \mathcal{L} = - 0.5 \cdot \left(\frac{f_\mathrm{ model} - f_\mathrm{ data}}{f_\mathrm{ error}}\right)^2 - 0.5 \cdot w \cdot \log_{10} \left(\frac{l_\mathrm{ don,model} - l_\mathrm{ don,data}}{\ln(10)}\right)^2 .
\end{equation}.

Here, $f_\mathrm{ model}$ is the mock photometry, while $f_\mathrm{ data}$ the observations in the various passbands, and $f_\mathrm{ error}$ their associated errors. $l_\mathrm{ don,model}$ is the ratio of $f_\mathrm{ donor}$ over $f_\mathrm{ total}$ in the simulated model fluxes, and $l_\mathrm{ don,data}$ the relative light contribution of the donor as a function of wavelength determined previously. The weight $w$ is chosen to ensure the flux-ratio term does not dominate the likelihood function.

Figure \ref{fig:SEDs} shows the results of the SED fitting, where we overplot the observed photometry (lime dots) with the SEDs corresponding to our spectroscopic temperatures and radius estimates, both for the whole binary system (black line) and for the individual components (magenta and cyan lines).  The bottom sub-panel for each object shows the spectroscopically determined donor flux contribution (red line), compared to that implied by the two components' SED.

\begin{figure*}
    \centering
    \includegraphics[width=1.0\linewidth]{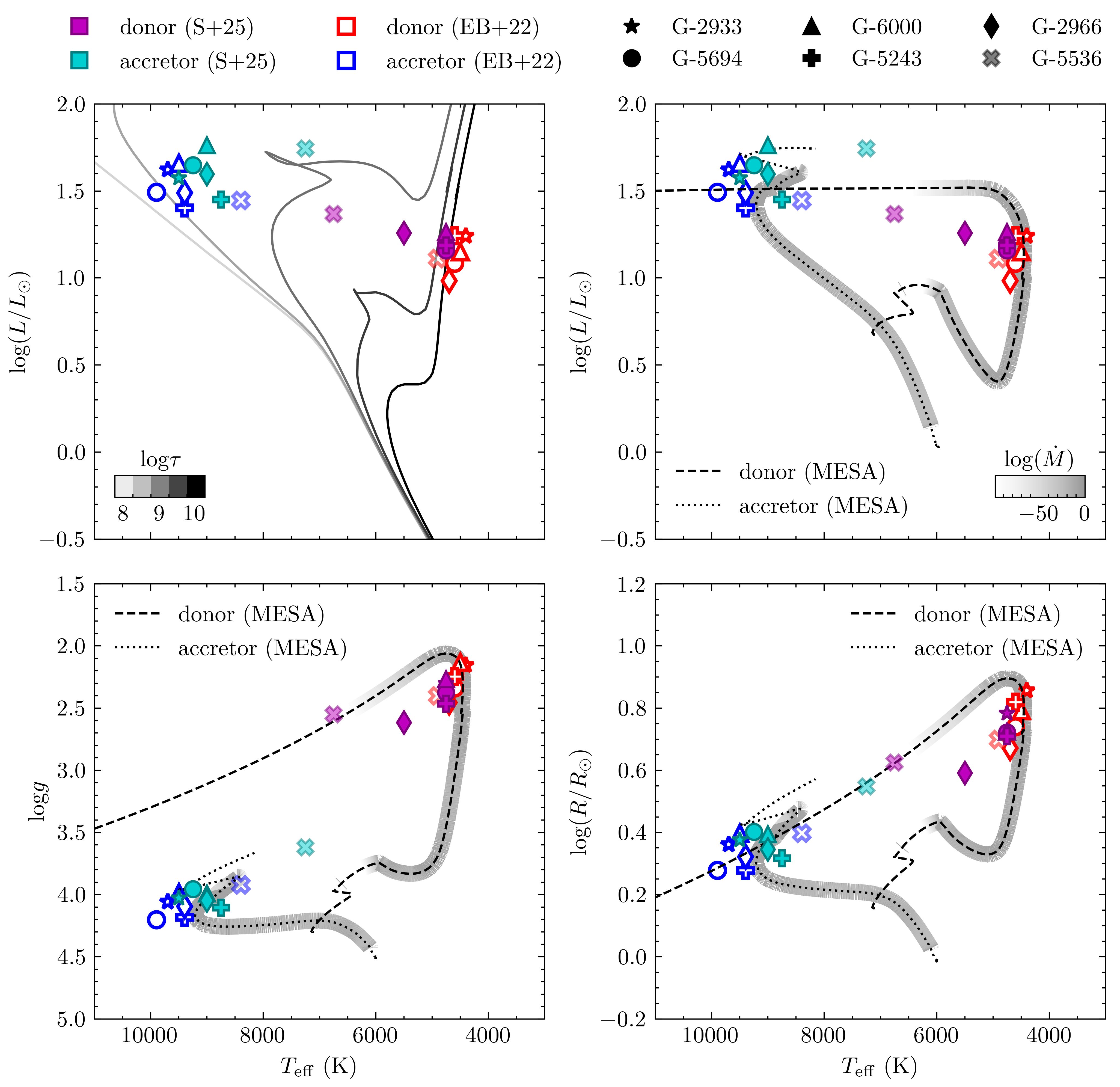}
    \caption{Parameters of both components of each system as determined in this work (S+25) and \citetalias{el-badry_what_2022}. Findings from S+25 are shown with filled symbols in magenta (donor) and cyan (accretor), while those from \citetalias{el-badry_what_2022} are empty, outlined in red (donor) and blue (accretor). Different symbols have been chosen to represent the six different systems, with the problematic system (G-5536) shown as more transparent than the rest. The top left panel shows a Hertzsprung-Russel diagram, including PARSEC isochrones spanning a range of ages \citep{bressan_parsec_2012a}. The top right panel shows the same parameter space, but with possible MESA  evolutionary paths for the donor (dashed line) and accretor (dotted line) included, as computed by \citetalias{el-badry_what_2022}. The bottom two panels show $\log{g}$ vs. $T_\mathrm{ eff}$ and $\log{R}$ vs $T_\mathrm{ eff}$ respectively, as well as including the aforementioned MESA tracks. The grey shading behind each line indicates the amount of mass transfer in the MESA model at that evolutionary stage.
    }
    \label{fig:HRD}
\end{figure*}

In Figure \ref{fig:HRD} (top panels) we summarise information from the spectral fits (effective temperatures) and SEDs (stellar radii) in a Hertzsprung-Russell diagram. In the top left panel, we also include PAdova and TRieste Stellar Evolution Code \citep[PARSEC, ][]{bressan_parsec_2012a} isochrones generated using the CMD 3.8 input form \footnote{https://stev.oapd.inaf.it/cgi-bin/cmd} for stellar logarithmic ages ranging from eight to ten in increments of 0.5, assuming solar metallicity.  In the remaining panels, we show the target parameters alongside Modules for Experiments in Stellar Astrophysics \citep[MESA, ][]{paxton_modules_2011a, paxton_modules_2013a, paxton_modules_2015, paxton_modules_2018a, paxton_modules_2019a, jermyn_modules_2023a} evolutionary tracks computed by \citetalias{el-badry_what_2022}. Here, the MESA tracks assumed a binary with main sequence (before mass transfer started) masses of 1.5 $M_\odot$ and 1.1 $M_\odot$ for the donor and accretor respectively. The initial period was assumed to be 1.3 days. Further details are described in \citetalias{el-badry_what_2022}.  
The positions of the components are also shown in the $\log{g}$-$T_\mathrm{ eff}$ and $\mathrm{log}(R)$-$T_\mathrm{ eff}$ planes (bottom panels). Here, $\log{g}$ was determined not from our disentangled spectra, but by combining the masses from \citetalias{el-badry_what_2022} with the radii from our SED fits, as surface gravities from the spectra proved to be poorly constrained.

\subsection{Orbital fitting}
\label{subsec:orbit}

Having determined reliable RVs for both components in Section \ref{subsec:disentangling}, we then used these to fit orbital solutions for each system. This step linked the spectroscopic results with the systems’ dynamical properties and provided the orbital periods required for the mass transfer analysis in Section \ref{subsec:MT_history}.

Similar to \cite{muller-horn_binary_2024}, we used a nested sampling framework to estimate posterior probability distributions for the orbital parameters. Nested sampling is a variation of the commonly used Markov chain Monte Carlo sampling algorithm to sample a posterior distribution \citep{skilling_nested_2004}. Here, we used the \verb|UltraNest| \footnote{https://johannesbuchner.github.io/UltraNest/} package \citep{buchner_ultranest_2021}, an implementation of the nested sampling algorithm.

The RV curve is described by six parameters: $(K, P, M_0, e, \omega, v_\mathrm{COM})$. $K$ represents the RV semi-amplitude of the visible star, $P$ is the orbital period, and $M_0 = \frac{2\pi t_0}{P}$  defines the mean anomaly at a reference time $t_0$. The parameters $e$ and $\omega$ correspond to the orbital eccentricity and argument of pericentre, while $v_\mathrm{COM}$ denotes the system's barycentric velocity. Uniform, non-informative prior distributions were adopted for all parameters, with prior ranges set as follows: $K \in [0,500]\,\text{km/s}$, $P \in [1,100]\,\text{d}$, $M_0, \omega \in [0,2\pi]$\, $e \in [0.0,0.9]$, and $v_\mathrm{COM} \in [-100,100]\,\text{km/s}$. The results for these parameters based on the RVs determined in this work are summarised in Table \ref{tab:orbital} in the Appendix.

\subsection{Mass transfer history}
\label{subsec:MT_history}

The MESA evolutionary tracks shown in Figure \ref{fig:HRD} were taken from \citetalias{el-badry_what_2022}. Detailed setup can be found there, but for the purpose of this work we want to highlight the adoption of a fully conservative mass transfer scheme, with mass transfer rates computed using \cite{kolb_comparative_1990a}. These assumptions successfully reproduced the observed locations of the components in the Hertzsprung–Russell diagram. In this section, we tested these assumptions by combining our updated orbital parameters and spectroscopic mass ratios to infer the possible mass transfer histories of our targets following the formalism of \cite{soberman_stability_1997a}.

Using the period from the orbital fit (c.f. with \citetalias{el-badry_what_2022} and Gaia in Table \ref{tab:orbital}), and the mass ratio from the disentangling (c.f. with \citetalias{el-badry_what_2022} in Table \ref{tab:params}), we placed constraints on how conservative mass transfer must have been in each system's past. We used equations from \cite{soberman_stability_1997a}, which relate the period and mass ratio in a mass transferring system subject to the mass transfer parameters $\alpha$, $\beta$ and $\delta$. Here, the parameters describe the fraction of the transferred mass lost via various channels.

$\alpha$ is the fraction lost via a wind from the donor, $\beta$ is the fraction ejected from the accretor, and $\delta$ is the fraction retained in a circumbinary ring. $\alpha$, $\beta$ and $\delta$ are defined as, respectively, 

\begin{align}
    \alpha = \frac{\partial m_\mathrm{wind}}{\partial m_\mathrm{don}}
    &&
    \beta = \frac{\partial m_\mathrm{iso-r}}{\partial m_\mathrm{don}}
    &&
    \delta = \frac{\partial m_\mathrm{ring}}{\partial m_\mathrm{don}} .
\end{align}

Here, $\partial m_\mathrm{wind}$ is the mass of stellar wind from the donor, $\partial m_\mathrm{iso-r}$ the mass of stellar wind from the accretor (isotropic re-emission) and $\partial m_\mathrm{ring}$ the mass used in the formation of a ring. $\partial m_\mathrm{don}$ is the mass lost from the donor. If all these parameters are set to zero, mass transfer is fully conservative, i.e. no mass is lost and 100\% of the mass donated by the donor is accreted by the accretor. If these parameters add up to one, then mass transfer is fully non-conservative, meaning all mass ejected by the donor is lost to the surrounding medium; the accretor does not successfully accrete any of it. Functionally, the relative values of $\alpha$, $\beta$, and $\delta$ determine the amount of angular momentum carried away by the matter lost from the system, and thus the extent to which the orbit widens or tightens.

For simplicity (following \cite{bodensteiner_hr_2020a,el-badry_unicorns_2022}) we assumed $\alpha = \delta = 0$, meaning no mass is lost via wind from the donor or to a circumbinary ring. This scenario is commonly referred to as isotropic re-emission. By varying $\beta$, we explored a number of more or less conservative mass transfer scenarios, particularly the evolution of the period with the changing mass ratio. Further, we set a minimum period $P_\mathrm{ min}$, which is where the donor overflows its Roche lobe on the main sequence \citep{eggleton_aproximations_1983}. The argument here is that we assume the binary was detached at some point in its evolution. If the mass transfer was too non-conservative (i.e. $\beta \gtrsim$0.7), the past orbit would have been too tight, and the donor would not have fit. For an in-depth analysis of binary contact tracing, see \cite{henneco_contact_2024a}.

\begin{figure}
    \centering
    \includegraphics[width=\linewidth]{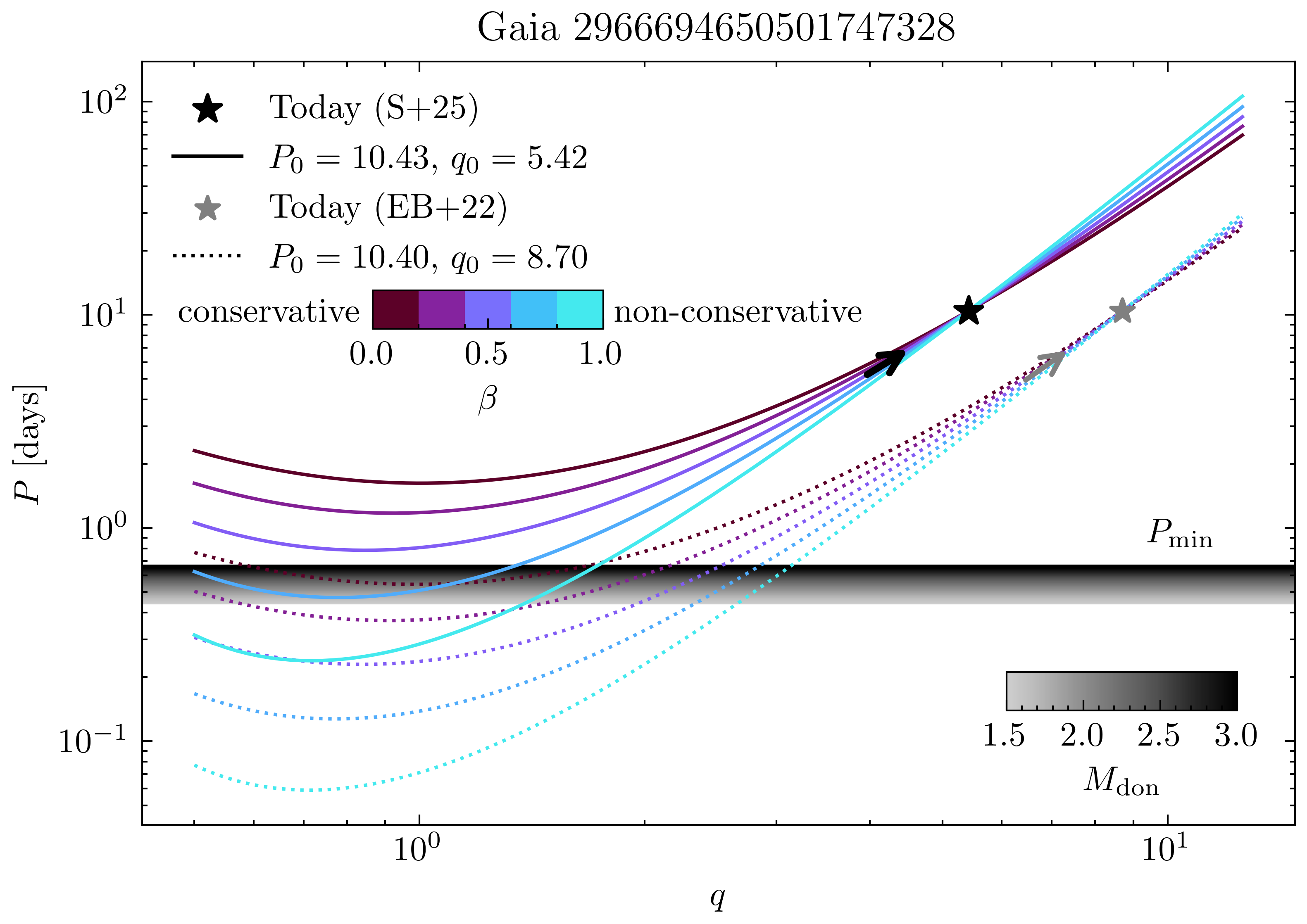}
    \caption{Period vs mass ratio for one of the targets (G-2966), showing the evolution of the period as mass is transferred from the donor to the accretor. Solid lines use present-day parameters found in this work (S+25), while dotted lines use the results from \citetalias{el-badry_what_2022}. The present-day locations of the systems in the P-q plane are marked with black (this work) and grey (\citetalias{el-badry_what_2022}) stars. Arrows indicate the direction of evolution as mass is transferred from the donor onto the accretor. The minimum Period is shown as a grey horizontal stripe, with colours indicating the donor mass on the main-sequence. We see that the higher mass ratio from \citetalias{el-badry_what_2022} places a much tighter constraint on the how conservative mass transfer was, than the lower estimates found in this work. For this object in particular, the present-day mass ratio and Period from \citetalias{el-badry_what_2022} require almost fully conservative mass transfer to avoid a merger.}
    \label{fig:masstransfer}
\end{figure}

Figure \ref{fig:masstransfer} shows the past evolution of the period as the system underwent mass transfer, for different $\beta$ values. At $q$ equal to one, the mass ratio is flipped, and the (initially) more massive donor is now the less massive star of the binary. The coloured lines show how the past period depends on the assumed $\beta$. Some, but not all, of these lines dip below the minimum period (grey gradient, with darker colours corresponding to higher, and lighter colours to lower main sequence masses). Below this minimum period, the radius of the donor's progenitor on the main sequence would have exceeded its Roche lobe, calculated using \cite{eggleton_aproximations_1983}. These reflect values of $\beta$ excluded by the data. 

In Figure \ref{fig:masstransfer}, the solid lines reflect the possible $P(q)$ for our determination of the present day mass ratio and period, while the dashed lines reflect the $P(q)$ implied by \citetalias{el-badry_what_2022}. This illustrates the importance of the $q$ determination, as it immediately translates into constraints on the mass transfer `physics', especially $\beta$. Analogous plots for the other systems are found in Figures \ref{fig:masstransfer_1} and \ref{fig:masstransfer_2} in the Appendix.

The inferred range of $\beta$ values is broadly consistent with the conservative mass transfer prescriptions assumed in the MESA models shown in Figure \ref{fig:HRD}. Consequently, our results do not require significant adjustments to the evolutionary tracks adopted by \citetalias{el-badry_what_2022}.

Our findings therefore support the \citetalias{el-badry_what_2022} conclusion that mass transfer in these systems must have been relatively conservative, with $\beta \lesssim 0.7$, as for larger $\beta$ (i.e. more mass loss), the period would be too short at some point during the binary's evolution. This affirms the conclusions of \cite{bodensteiner_hr_2020a}, \citetalias{el-badry_what_2022} and \cite{lechien_binary_2025a} that mass transfer has to be quite conservative, and is consistent with the MESA evolutionary paths in Figure \ref{fig:HRD}.

\section{Discussion}
\label{sec:discussion}

Having derived the key stellar, orbital, and flux parameters, we now discuss their implications for the current evolutionary state of these binaries and for the physics of mass transfer and angular momentum evolution. Taken together, our spectroscopic analysis showed and confirmed that all these systems have all expected signatures of post-mass-transfer binaries. Spectral disentangling revealed a cool, synchronously rotating donor and a hot, rapidly rotating accretor. The systems show no emission lines (except for object G-2933), indicating that there is no significant mass transfer at present. 

More quantitatively, our analysis provided four new pieces of information on each system. These are a measure of the temperatures, $T_\mathrm{ eff}$, using the two best-fitting templates, independent from those computed by \citetalias{el-badry_what_2022}; the wavelength-dependent flux ratio of both components; their projected rotation velocities, $v\sin{i}$; and the kinematically determined mass ratio, $q$.

We illustrate the first two aspects in Figure \ref{fig:SEDs}. The SEDs with the parameters inferred by \citetalias{el-badry_what_2022} are shown as dotted lines in the top sub-panels for each object, while those using parameters determined in this work are shown as solid lines. Comparing our spectroscopic constraints with the SED constraints from  \citetalias{el-badry_what_2022} shows broad consistency.
For all objects, we find spectroscopic donor temperatures that are slightly higher than those in \citetalias{el-badry_what_2022}. Similarly, four of the six accretors also show higher temperatures than those found by \citetalias{el-badry_what_2022}.
It must be noted, however, that the accretor temperatures lie in the regime of T$_\mathrm{eff}\approx 9000$ K. Here, spectral modelling assuming local thermal equilibrium \citep{coelho_new_2014, xiang_stellar_2022}, including ours, has difficulty matching the spectra, especially the very strong Balmer lines. The net effect of the smaller temperature difference among the components is a more gradual change in the component flux ratio with wavelength in our results, compared to \citetalias{el-badry_what_2022}.

Figure \ref{fig:SEDs} shows that the SEDs predicted by our spectroscopic temperatures match the observed photometry almost as well as the SED fits in \citetalias{el-badry_what_2022}, after finding the best fitting stellar radii. This presumably reflects the uncertainties and covariances inherent in multiparameter photometry fits, which can be mitigated with our spectroscopic temperatures and component flux ratio estimates.  

The Hertzsprung-Russell diagram in Figure~\ref{fig:HRD} (top panels) illustrates the nature of the systems, as well as the similarities and differences among them. We show our results (magenta and cyan filled symbols for the donor and accretor, respectively) and compare them to those from \citetalias{el-badry_what_2022} (red and blue empty symbols for the donor and accretor, respectively). For context, we show single-star PARSEC isochrones \citep{bressan_parsec_2012a} of various ages in the top left panel and pertinent MESA \citep{paxton_modules_2011a, paxton_modules_2013a, paxton_modules_2015, paxton_modules_2018a, paxton_modules_2019a, jermyn_modules_2023a} stellar evolution tracks for the donor and accretor computed by \citetalias{el-badry_what_2022} in the top right. Analogously, we show the positions of all components in the $\log{g}$-$T_\mathrm{ eff}$ and $\log{R}$-$T_\mathrm{ eff}$ planes (bottom panels), along with the aforementioned MESA tracks.

Overall, we see qualitatively good agreement between the two works, with the exception of object G-5536. This is reassuring, as the parameters for the systems were determined via fundamentally different pathways, increasing our confidence in the congruent results. Small discrepancies in the determined effective temperatures of the stellar component may be attributed to the degeneracies in the SED fitting process employed by \citetalias{el-badry_what_2022}, which are lifted in this work using spectroscopic temperatures from disentangled spectra.

We also note that the accretors are consistent with much younger (single-star) isochrones than the donors, an expected outcome of mass transfer onto the accretor and the resulting rejuvenation. Both sets of results are also roughly consistent with the fiducial MESA tracks from \citetalias{el-badry_what_2022}, which we did not try to optimise for all the individual systems.

Note that our sample contains one system where all these analysis steps do not work as well (G-5536), starting with a poor spectral disentangling result (see Figure~\ref{fig:initspectra}, which we attribute -- at least in part -- to the observations' lower signal-to-noise ratio. This is also encapsulated in the less good fit of the model flux ratio (bottom panel, black line) to the observed spectroscopic flux ratio (red line).

We originally anticipated that our high-resolution multi-epoch spectra should enable tight constraints on the systems' mass ratio, $q$. But using TODCOR \citep{zucker_study_1994} with our best-fitting templates still faced serious difficulties, manifested, for example, as inconsistent velocity estimates when using different wavelength windows to derive them. We attribute this to several compounding factors. First, the accretors are rapidly rotating, which severely broadens their lines, and have few strong lines beyond the inherently broad Balmer series. Second, the spectra in some of our systems (especially  object G-2933, showing an emission signature) are unusual. This may lead to rather imperfect template match and in turn may cause systematic problems in TODCOR. This is exacerbated by the fact that the five to ten times more massive accretor has both far smaller velocity variations, and no sharp spectral lines. As the accretor contributes more at the shortest wavelengths, we deemed the $q$ estimates based on H$\gamma$ and H$\delta$ more reliable, compared to H$\beta$. We also explored whether the \ion{Mg}{i} and \ion{Mg}{ii} lines present in the accretor spectrum worked better, but found that this was not the case.

For most objects, our kinematic mass ratios are somewhat smaller than those in \citetalias{el-badry_what_2022}, consistent with our smaller temperature difference between the donor and accretor components. Considering Figure \ref{fig:TODCOR}, the significant scatter in the secondary velocities and resulting large uncertainty in the mass ratio (and systemic velocity) might call this result into question relative to \citetalias{el-badry_what_2022}. Indeed, we have found the determination of the mass ratio to be very sensitive to the templates chosen for both components, with lower-temperature donor templates leading to larger implied dynamical mass ratios. Best-fit templates are determined in a specific wavelength region, which is distinct from the windows used here with TODCOR. As such, a different template might produce a better fit in the window in question, but we found it more important to use the same template consistently for all wavelength windows.

Despite some issues with the template match and the difficulties due to broad accretor lines, we point out that this work represents a much more direct measurement of the mass ratio, which has the distinct advantage of not requiring assumptions about the Roche Lobe filling factor or orbital inclination of the system, as in \citetalias{el-badry_what_2022}. Future work using spectral templates that incorporate a scheme that does not assume local thermal equilibrium may lead to more accurate and reliable results, especially considering the effective temperatures of the accretors.

The determination of the mass ratio has a large knock-on effect on the constraints on how conservative mass transfer was (the $\beta$ parameter in the \cite{soberman_stability_1997a} equations). In Figure \ref{fig:masstransfer}, we see that the updated mass ratio in this work relative to \citetalias{el-badry_what_2022} leads to different constraints on the maximum fraction of mass lost by the accretor. Generally, the lower mass ratios found in this work lead to more relaxed constraints. As these constraints represent upper limits on the mass loss, our results are consistent with the tighter limits found by using parameters from \citetalias{el-badry_what_2022}. In a few cases (G-5694, G-5536), parameters from \citetalias{el-badry_what_2022} imply mass transfer histories that are inconsistent with anything but almost fully conservative mass transfer ($\beta \lesssim0.2$), and for G-6004, the modern period and mass ratio determined by \citetalias{el-badry_what_2022} are inconsistent with the assumption of a system which was detached in the past, even with fully conservative mass transfer (see Figures \ref{fig:masstransfer_1} and \ref{fig:masstransfer_2} in the appendix).

For object G-5536, the mass ratio found here is close to one (depending on the wavelength range used). However, as can be seen in the top right panel of Figure \ref{fig:TODCOR}, the quality of the individual straight-line fits (which give mass ratio, $q$ and systemic velocity, $v_\mathrm{COM}$, see bottom right panel of Figure \ref{fig:TODCOR}), is not very good. This can be explained by reconsidering the disentangled spectra (Figure \ref{fig:initspectra}): the disentangled components for object G-5536 are noisy, making selecting the correct template difficult. This is partially due to the fact that object G-5536 is fainter than the rest of the sample, but was observed for a similar amount of total time, leading to a lower S/N overall. Additionally, object G-5536 is likely further evolved, since the independent template fits suggest a higher effective temperature for the donor than found in \citetalias{el-badry_what_2022}. The puffed-up stripped star has likely started contracting and heating up again. This can also be seen in Figure \ref{fig:HRD}: Specifically, we note that the G-5536 components lie further along their respective MESA evolution tracks in all three panels that include them. 

This further evolution of object G-5536 results in more similar temperatures of the donor and accretor, and thus a more similar spectral signature. Especially in the Balmer series, the higher donor temperature leads to wider and deeper lines, more similar to those of the accretor. This causes difficulties in the disentangling and subsequent TODCOR analysis.

\paragraph{Stellar rotation}

As a final point of discussion, we now turn to the level of rotation in the accretors that we determined from our disentangled spectra. These provide additional insights into the angular momentum evolution during and after mass transfer.

We see projected rotational velocities between 150 $\sim$ 200 km/s (see Table~\ref{tab:params}), a novel measurement not performed by \citetalias{el-badry_what_2022}. While this is clearly rapid rotation, as expected for an accretor, these velocities are well below the critical velocities that are expected to be $\gtrsim 350$~km/s (from $v_{crit} = \frac{G \cdot M_*}{R_*}$) for these systems, assuming masses and radii from \citetalias{el-badry_what_2022}. Following \cite{packet_spinup_1981a}, \cite{ghodla_evaluating_2023a} and \cite{henneco_contact_2024a}, we expect a star to have to accrete only about two to ten per cent of its initial mass to spin up to critical rotation. As our analysis suggests the accretors approximately doubled their mass through mass transfer, they should have easily achieved critical rotation.

This discrepancy between expected and observed rotational velocities is unlikely due to mere projection effects; \citetalias{el-badry_what_2022} found the inclinations of the systems to lie mostly around 65°, which would reduce the velocities only by about ten per cent. Significantly larger inclinations are ruled out as they would produce eclipses, while significantly smaller inclinations are inconsistent with observed accretor parameters. Part of this disagreement may stem from equatorial gravity darkening: due to the reduced gravity at the equator and the Von Zeipel effect \citep{vonzeipel_radiative_1924}, there is less flux coming from the equator, and the star appears to rotate less rapidly \citep{townsend_bestar_2004a}. However, as the observed discrepancy between observed and expected rotational velocity is large (of order factor two), it is likely that some mechanism has slowed the accretor's rotation. Mass transfer likely ended only recently, as the donors are still in the short-lived `puffed-up' stage \citep[see also][]{bodensteiner_hr_2020a, villasenor_btype_2023}. This implies that the spin-down mechanism must have been fairly efficient. Tides are likely in effect, as demonstrated by the donor rotating at tidally synchronous velocities and suggested by the short periods. However, tidal synchronisation following \cite{zahn_tidal_2008a} is significantly less efficient than required to reproduce the observed rotational velocities. Assuming parameters of a star representative of the accretors in our sample, we expect the tidal synchronisation timescale, $t_\mathrm{sync} \approx 10^{10}$ years. As we know, mass transfer only ceased recently (evidenced by the donors still being in the short-lived bloated stage) it is unlikely that tidal forces alone were sufficient to slow down the rotation of the accretor. Further, the accretors are too hot for magnetic braking \citep{kraft_studies_1967}. If star-accretion disk interactions, such as described by \cite{popham_does_1991a} and \cite{paczynski_polytropic_1991a}, are in effect, it is possible that these are, or were, slowing down the accretor. The mechanism allows for outward angular momentum transport at the same time as inward mass flow onto the star. These interactions may pose an alternative spin-up mechanism, as the conventional picture suggesting spin up to critical and then cessation of accretion leads to relatively low mass transfer efficiencies \citep[e.g.][]{henneco_contact_2024a}. 

\section{Conclusions}

In this work we have obtained and analysed multi-epoch spectra of six targets, identified by \citetalias{el-badry_what_2022} as likely post-mass-transfer binaries with a puffed-up donor and hot, rapidly rotating accretor. Our data and analysis yield component temperatures and sizes, and mass ratios for all systems, confirming this initial assessment. Only for one of the systems (G-2933) we found clear H$\beta$ emission, suggesting that mass transfer is still ongoing.

To better understand the physical state of the components of each system, we used the spectral disentangling approach by \citet{seeburger_autonomous_2024} to determine the effective temperatures and rotational velocities from the spectral features of each component. We found slightly higher temperatures (see Table \ref{tab:params}) for the donors than \citetalias{el-badry_what_2022} determined from SED fits. SED fits with the spectroscopic temperatures are still consistent with the observed photometry. 

We determined the dynamical mass ratios (Table \ref{tab:params}) of the targets using TODCOR \citep{zucker_study_1994}, finding somewhat lower mass ratios than \citetalias{el-badry_what_2022}. This is consistent with our finding of warmer donors, as warmer donors are likely more massive, reducing $q$. Combining the current orbital periods and mass ratios, we assessed the mass transfer history, following \cite{soberman_stability_1997a}, and found that the mass transfer had to have been fairly conservative ($\beta \lesssim 0.7$) to produce the observed systems. 

Finally, we determined that all accretors are rotating rapidly, but -- importantly -- considerably more slowly than the critical rotation velocity. This implies some mechanism must have slowed down rotation efficiently after mass transfer halted. Magnetic braking is unlikely to have been effective to slow down rotation, given that the $T_\mathrm{eff}\sim 9, 000$~K of the accretors is well above the `Kraft-break' \citep{kraft_studies_1967}. Given that we know the radii, masses and the orbit of the stars, the tidal spin-down prescriptions developed by \citet{zahn_tidal_2008a} imply slow-down timescales far in excess of the system ages. 

One target (G-5536) was consistently difficult to analyse, due to comparatively lower signal-to-noise data and higher similarity of the donor and accretor spectra and subsequently more degeneracies in the disentangling. For this target, additional epochs at higher signal-to-noise ratios might help shed more light on its nature.

\begin{acknowledgements}
RS, HWR, JMH and JV acknowledge support from the European Research Council through ERC Advanced Grant No. 101054731. JH is grateful for support from UK Research and Innovation (UKRI) in the form of a Frontier Research grant under the UK government’s ERC Horizon Europe funding guarantee (SYMPHONY; PI Bowman, grant number: EP/Y031059/1) and the European Research Council (ERC) under the European Union’s Horizon 2020 research and innovation programme (Starting Grant agreement N$^\circ$ 945806: TEL-STARS).

We thank Tomer Shenar for helpful discussion and input on this work.

The authors also would like to thank the anonymous referee for constructive and insightful comments.

This publication made extensive use of the online authoring Overleaf platform (\url{https://www.overleaf.com/}).\\

The data processing and analysis made use of
        \verb|pystellib| (Fouesneau),
        \verb|matplotlib| \citep{hunter_matplotlib_2007a},
        \verb|NumPy| \citep{harris_array_2020a},
        the \verb|IPython| package \citep{perez_ipython_2007},
        \verb|SciPy| \citep{virtanen_scipy_2020a},
        \verb|AstroPy| \citep{theastropycollaboration_astropy_2013, theastropycollaboration_astropy_2018, theastropycollaboration_astropy_2022}
        \verb|SpectRes| \citep{carnall_spectres_2017}
        and \verb|SpectResC| \citep{lam_spectres_2024}
\end{acknowledgements}

\bibliographystyle{aa}
\bibliography{library}

\begin{appendix}

\section{Disentangled spectra}

\begin{figure}
    \centering
    \includegraphics[width=\linewidth]{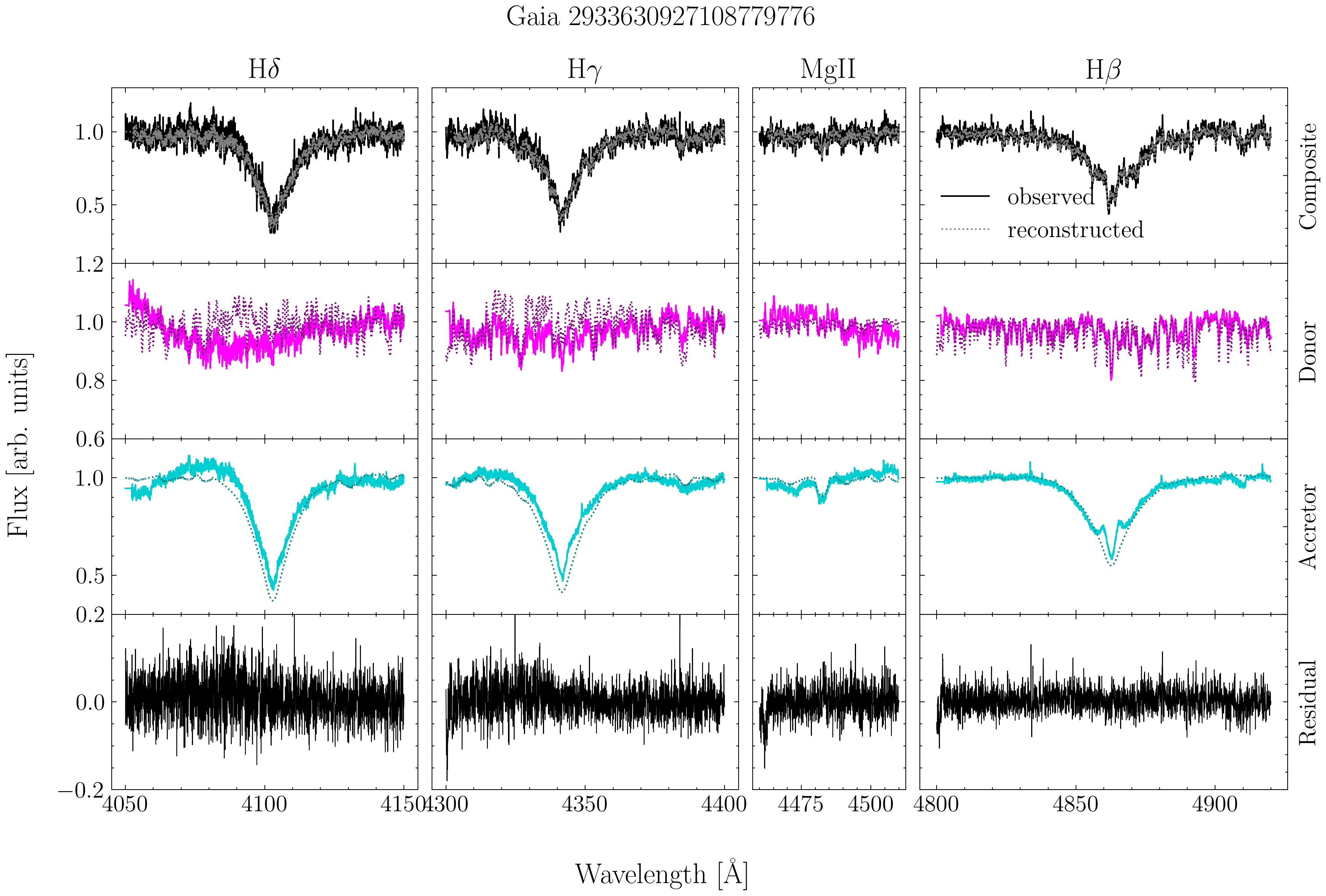}
    \includegraphics[width=\linewidth]{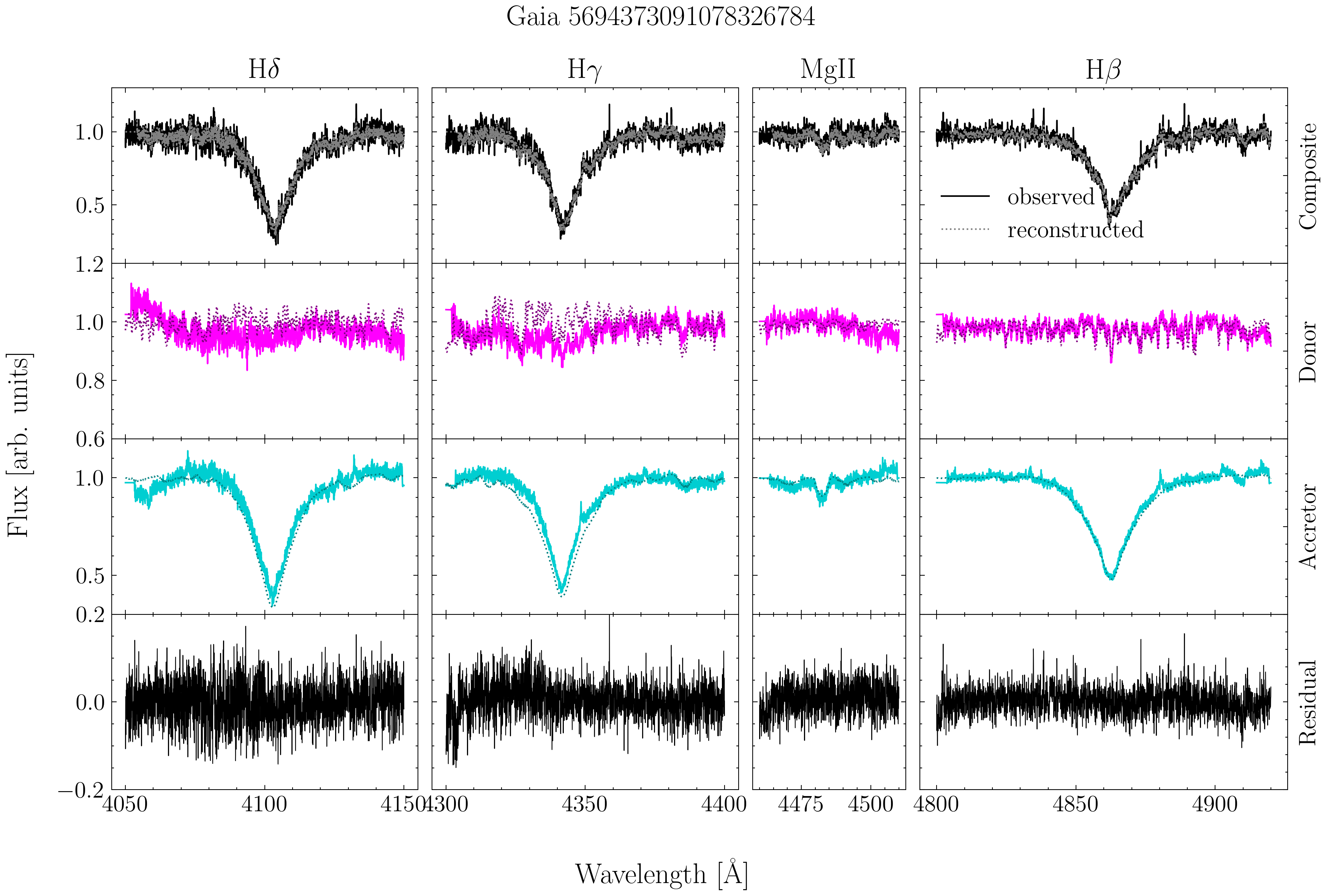}
    \includegraphics[width=\linewidth]{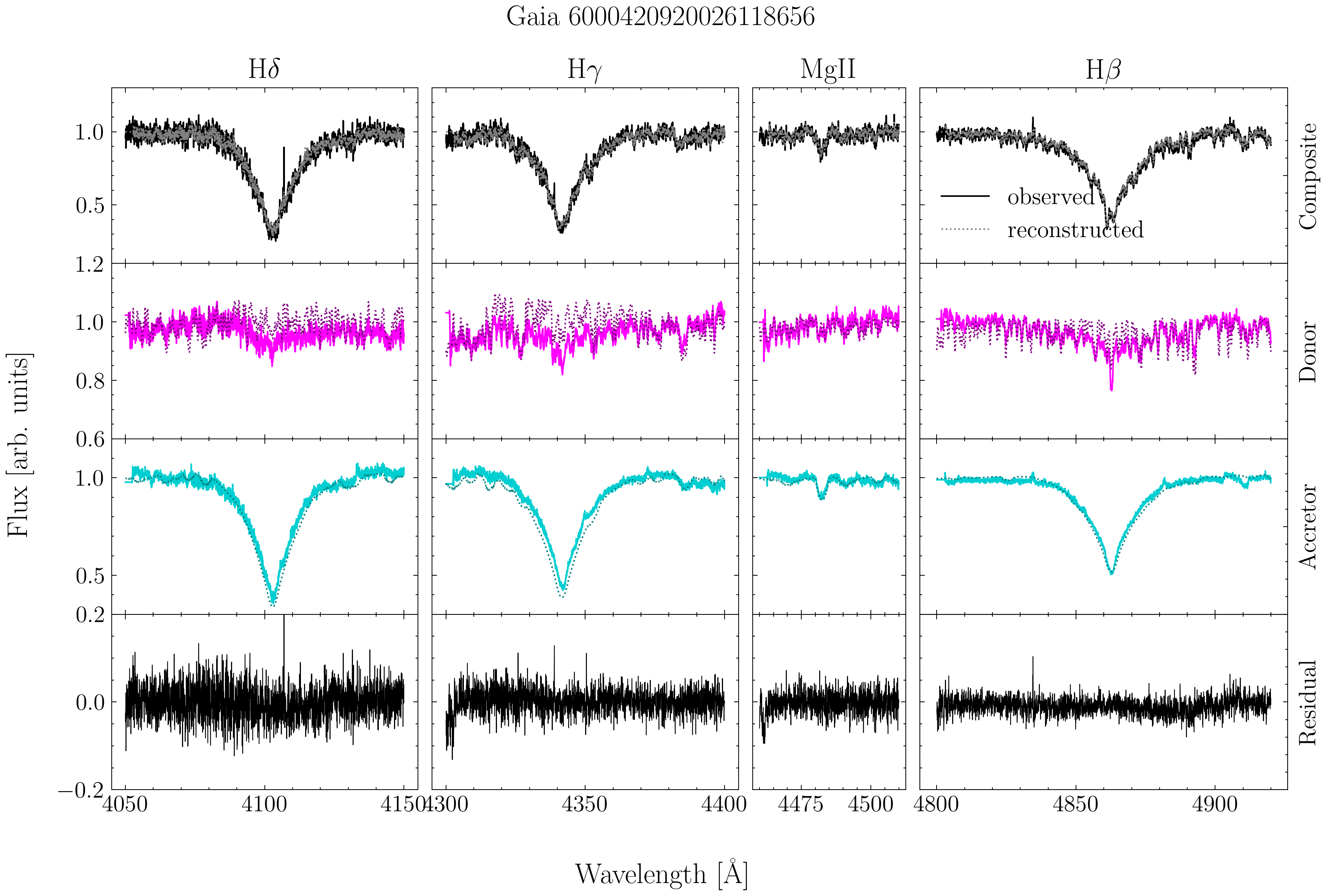}
    \caption{Disentangled spectra in different wavelength regions for three of the targets considered in this work, analogous to Figure \ref{fig:regions}. This demonstrates that one target, G-2933, does indeed show wings on either side of the H$\beta$ line, indicative of H$\beta$ emission and suggesting ongoing mass transfer.}
    \label{fig:regions1}
\end{figure}

\begin{figure}
    \centering
    \includegraphics[width=\linewidth]{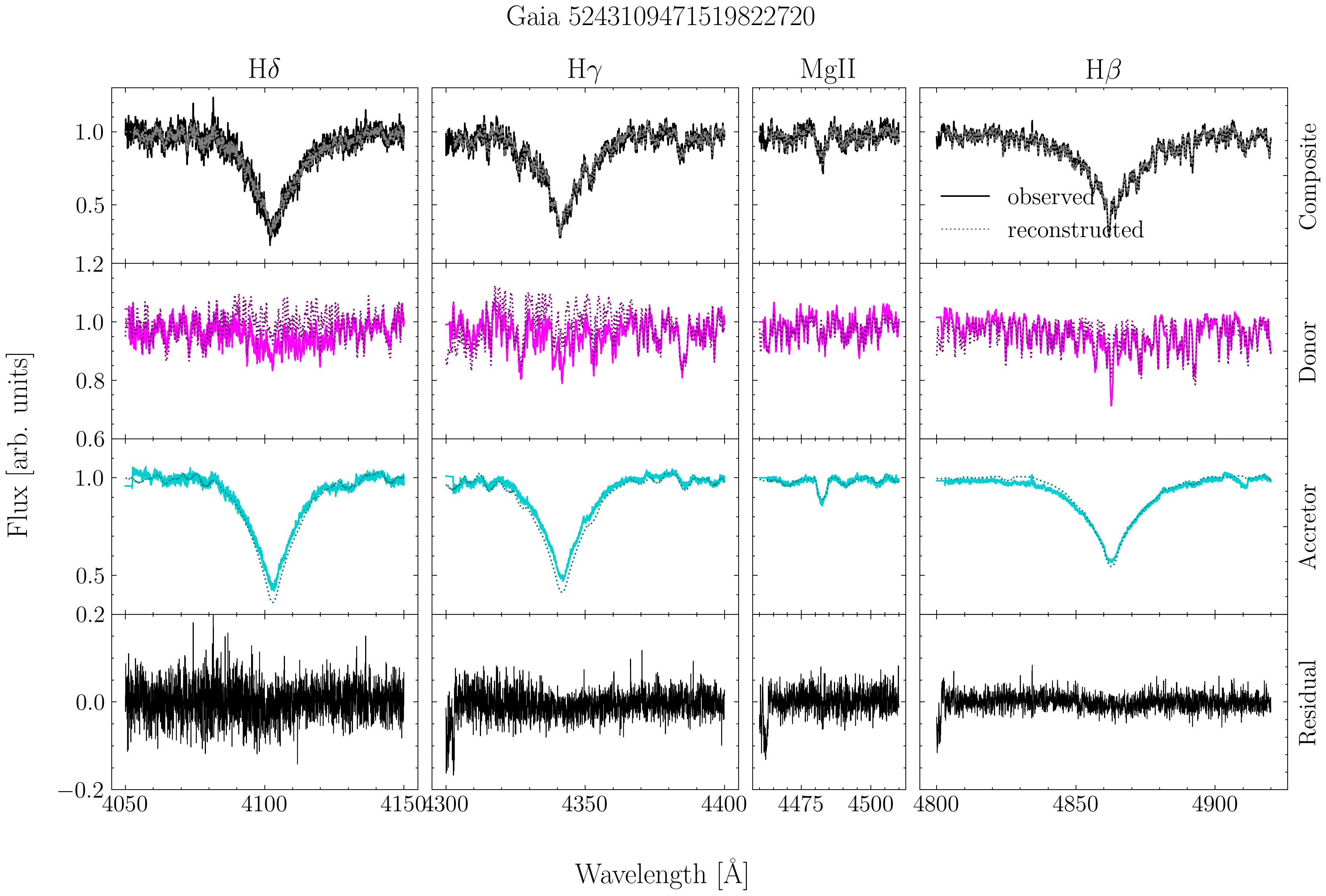}
    \includegraphics[width=\linewidth]{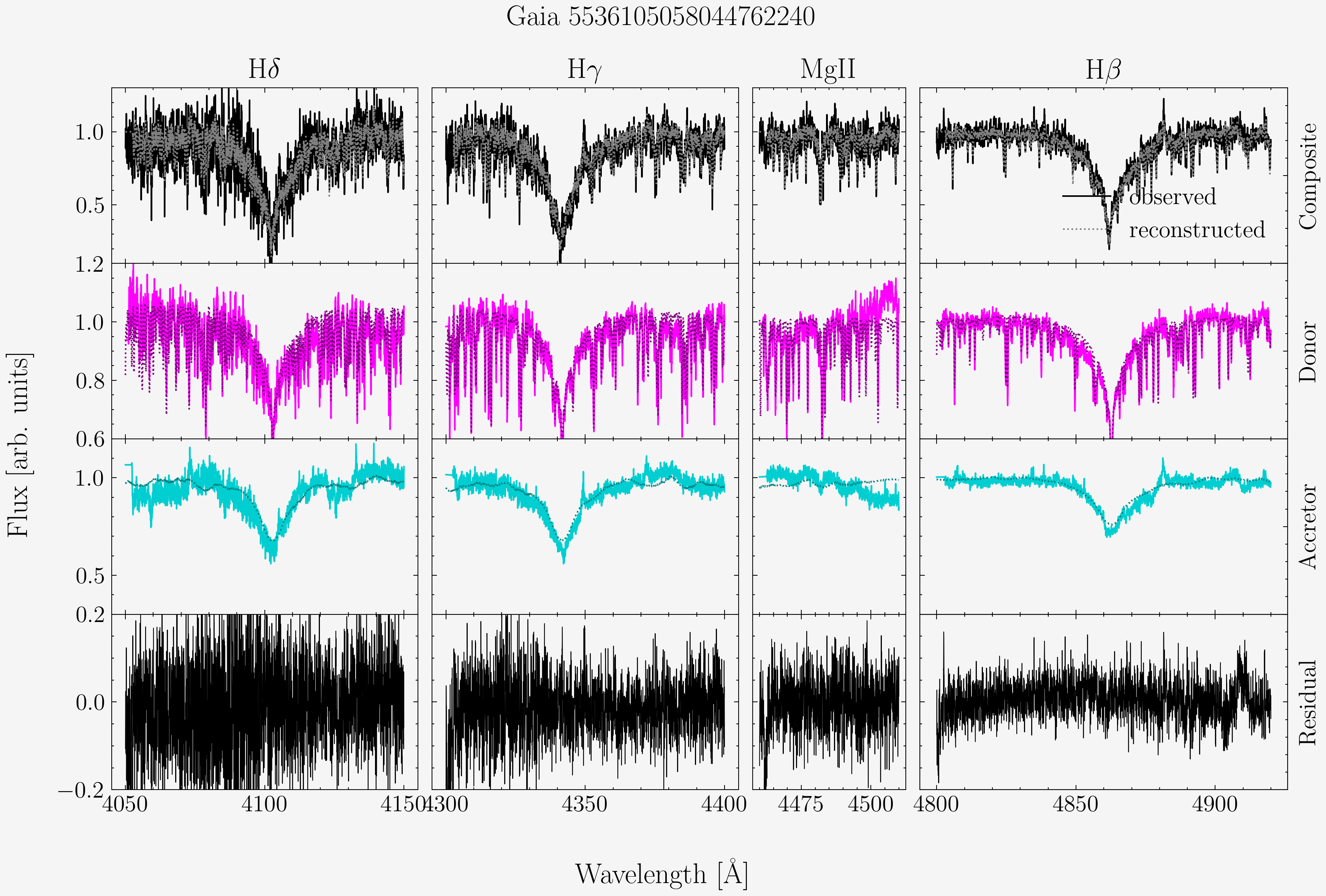}
    \caption{Disentangled spectra in different wavelength regions for two of the targets considered in this work, analogous to Figure \ref{fig:regions}. The background for the panel showing target G-5536 is shown in grey to highlight the difficulties with its analysis.}
    \label{fig:regions2}
\end{figure}

Figures \ref{fig:regions1} and \ref{fig:regions2} show the disentangled spectra for the remaining targets not displayed in the main text. Each panel illustrates the observed composite spectrum, the separated donor and accretor spectra in their rest frames, the best-fitting template spectra, and the residuals.
These figures complement Figures \ref{fig:initspectra} and \ref{fig:regions}, confirming that all six systems are double-lined spectroscopic binaries with two luminous components and demonstrating the quality of the disentangling across several wavelength regions.

\FloatBarrier
\section{Additional TODCOR results}

\begin{figure}[!htb]
    \centering
    \includegraphics[width=\linewidth]{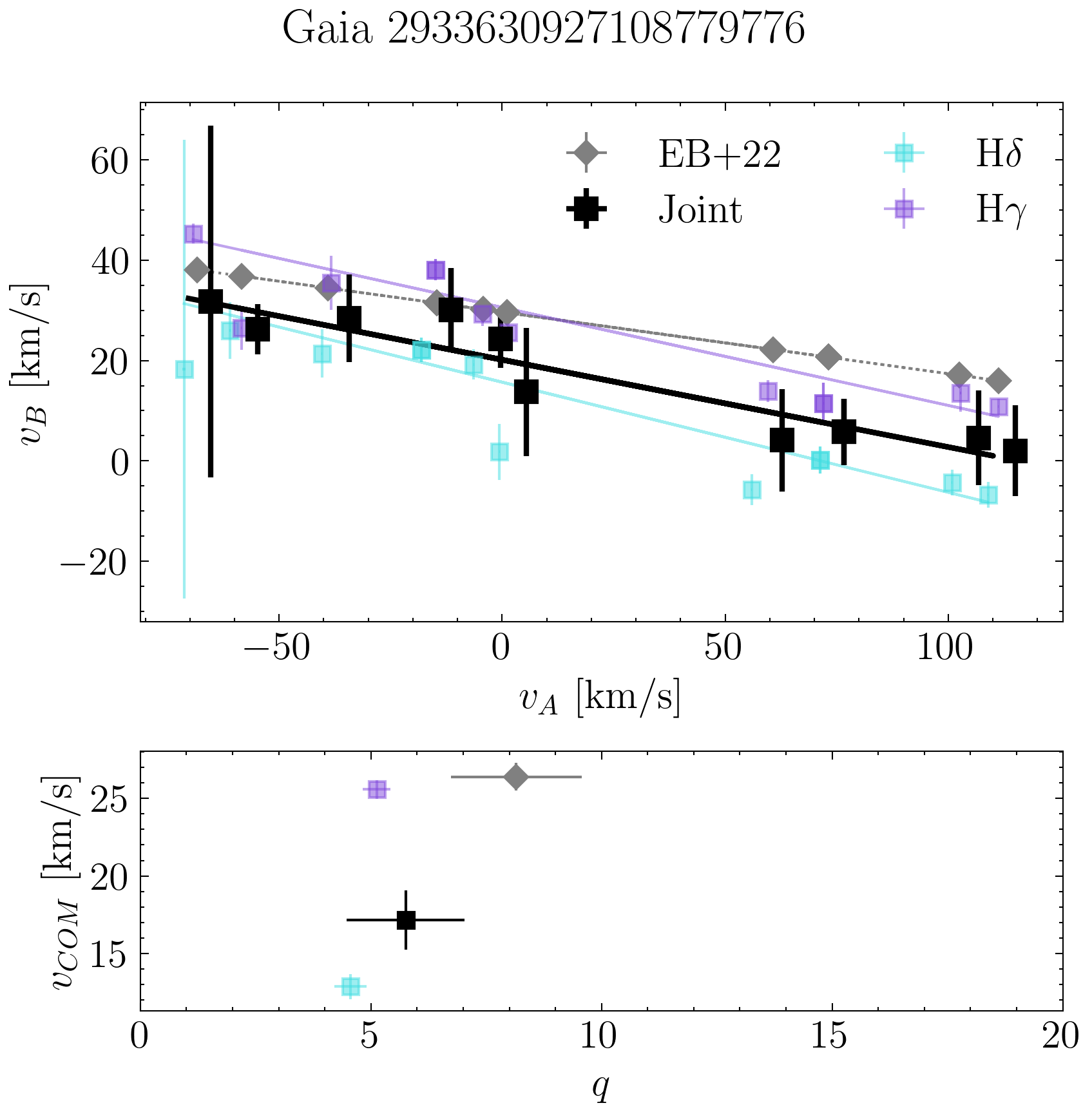}
    \includegraphics[width=\linewidth]{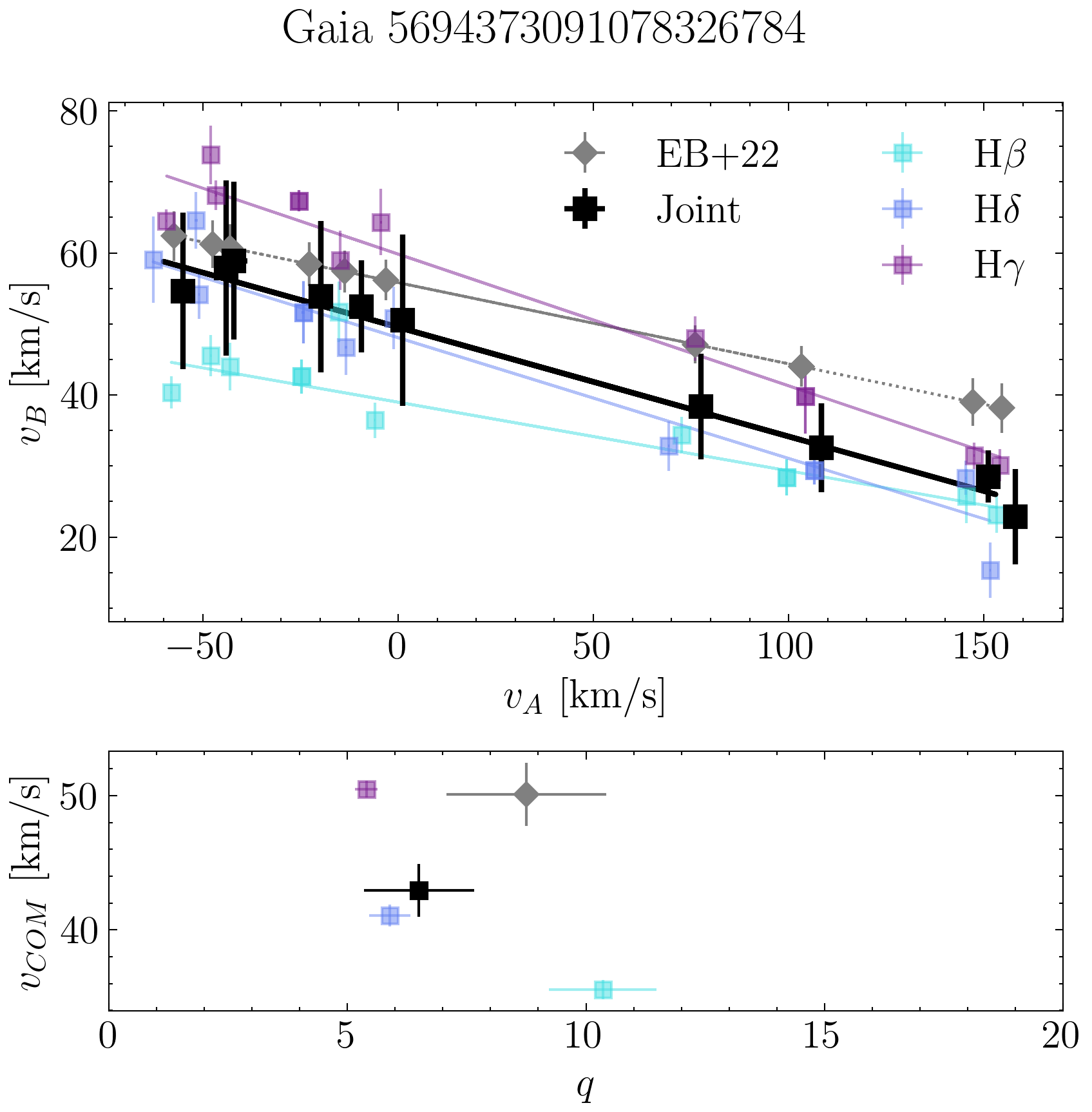}
    
    \caption{Results of the TODCOR \citep{zucker_study_1994} algorithm applied to two of the targets considered in this work, analogous to Figure \ref{fig:TODCOR}}
    \label{fig:TODCOR_1}
\end{figure}

\begin{figure}[!htb]
    \centering
    \includegraphics[width=\linewidth]{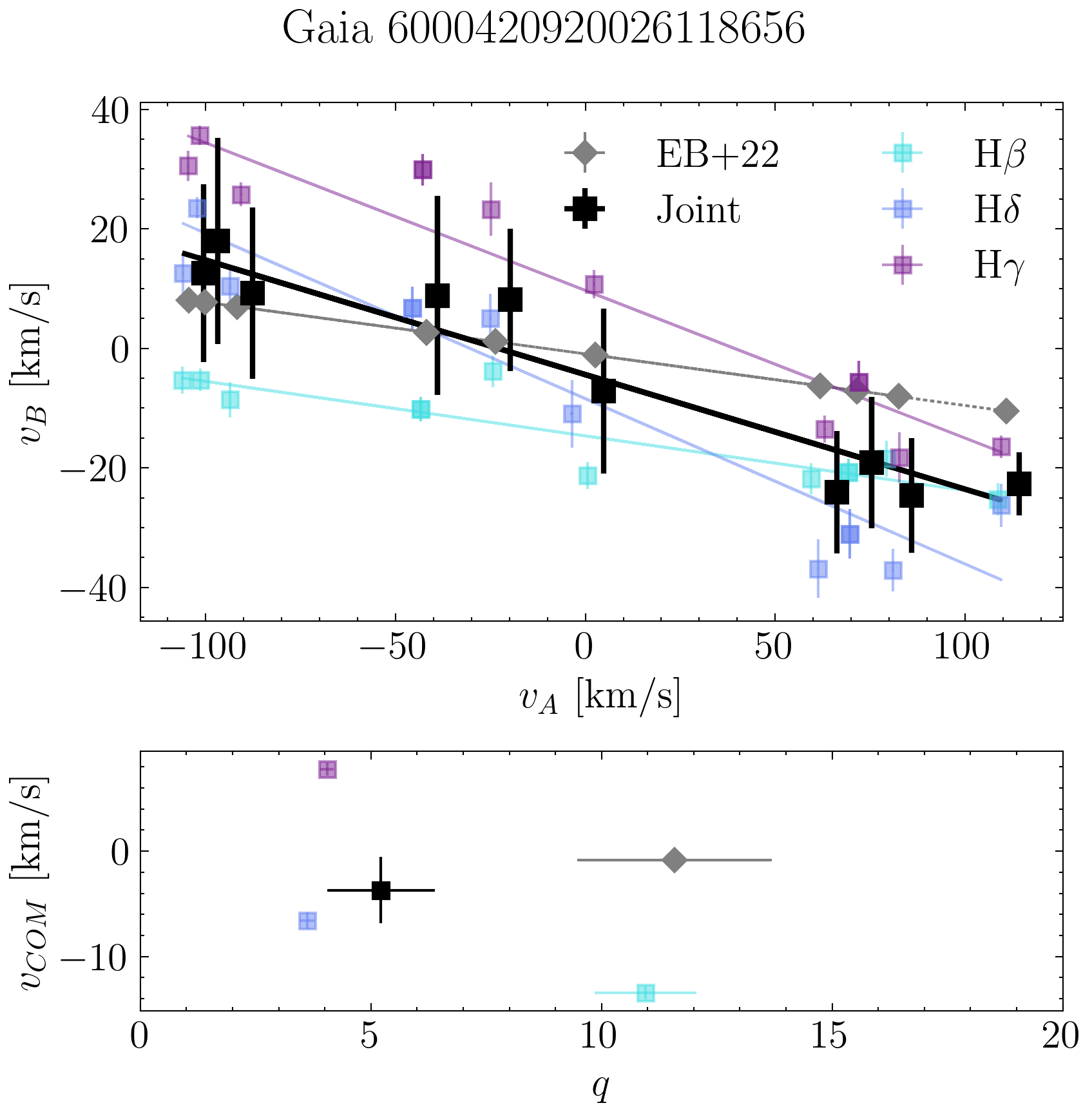}
    \includegraphics[width=\linewidth]{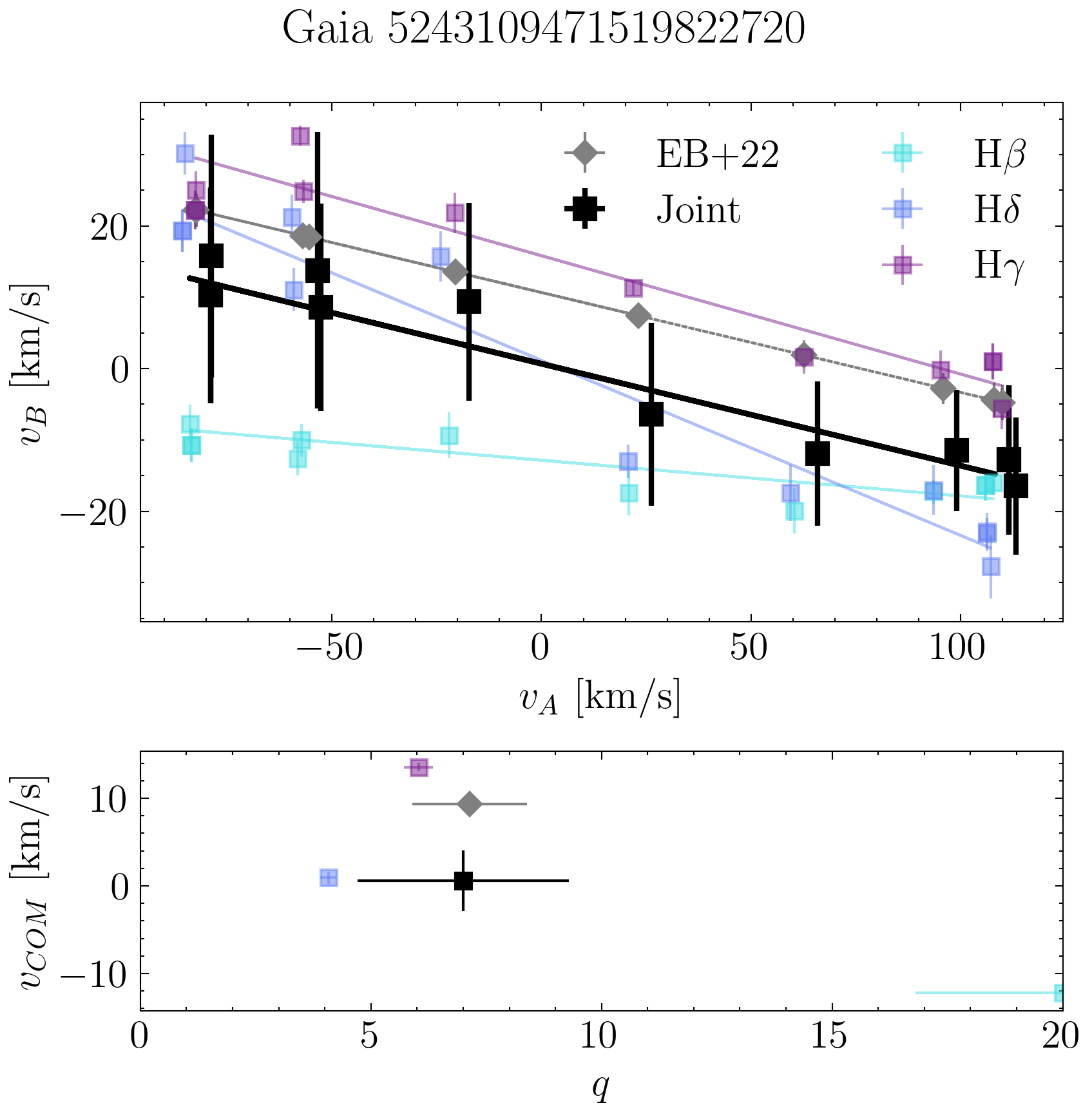}
    
    \caption{Results of the TODCOR \citep{zucker_study_1994} algorithm applied to two of the targets considered in this work, analogous to Figure \ref{fig:TODCOR}}
    \label{fig:TODCOR_2}
\end{figure}

This appendix presents the results of the TODCOR analysis described in Section \ref{subsec:TODCOR} for the remaining targets not shown in the main text.
For each binary, Figures \ref{fig:TODCOR_1} and \ref{fig:TODCOR_2} show the velocity–velocity diagrams of the donor and accretor components across multiple epochs and wavelength windows, together with the best-fitting linear relations used to determine the mass ratio and systemic velocity.
The lower panels display the corresponding distributions of derived mass ratios, $q$; and centre-of-mass velocities, $v_\mathrm{ COM}$; from individual wavelength intervals, compared to fiducial results from Gaia and \citetalias{el-badry_what_2022}.
These plots illustrate both the consistency and scatter of the TODCOR solutions.

\FloatBarrier
\section{Mass transfer histories}

\begin{figure}[!htb]
    \centering
    \includegraphics[width=\linewidth]{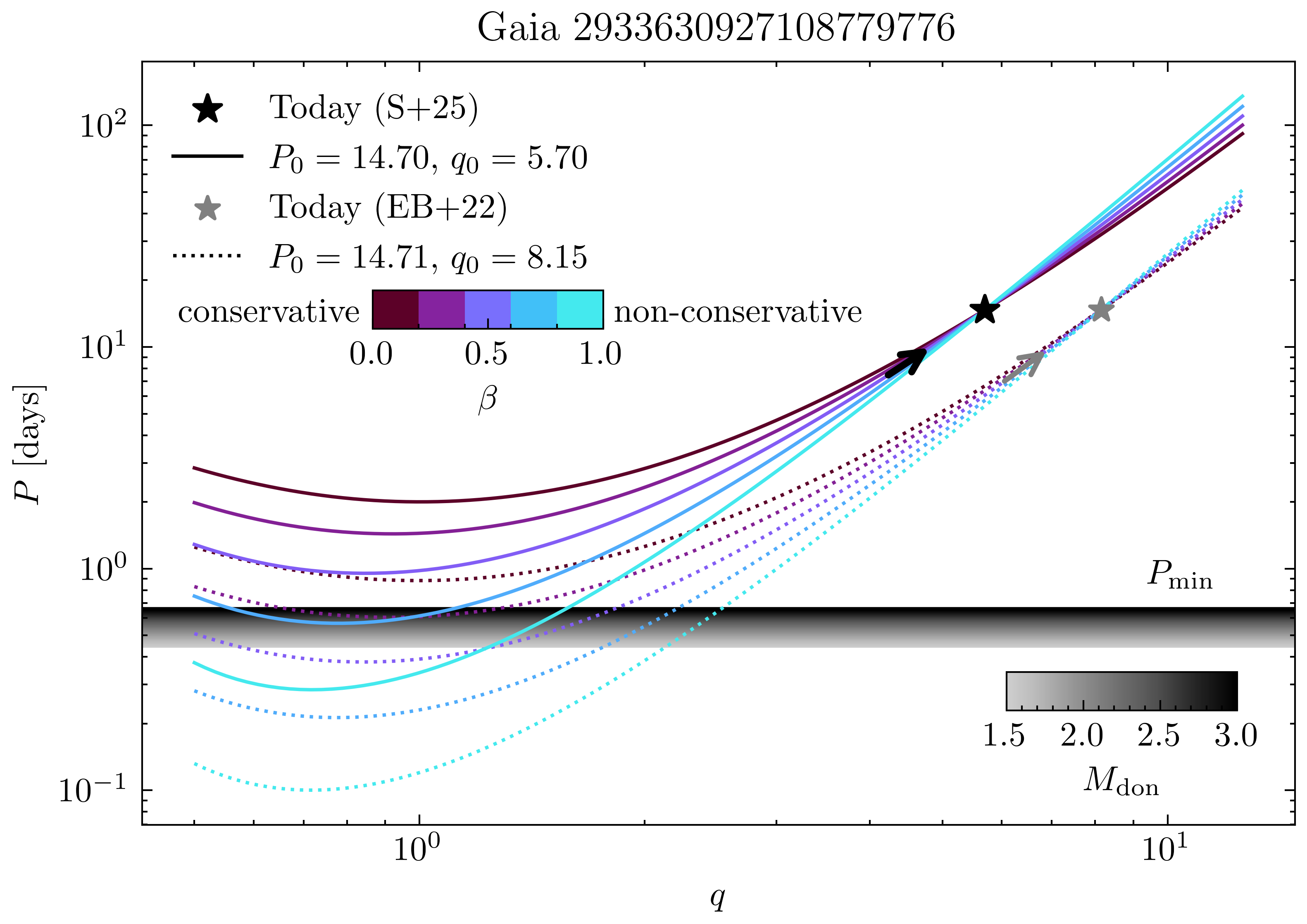}
    \includegraphics[width=\linewidth]{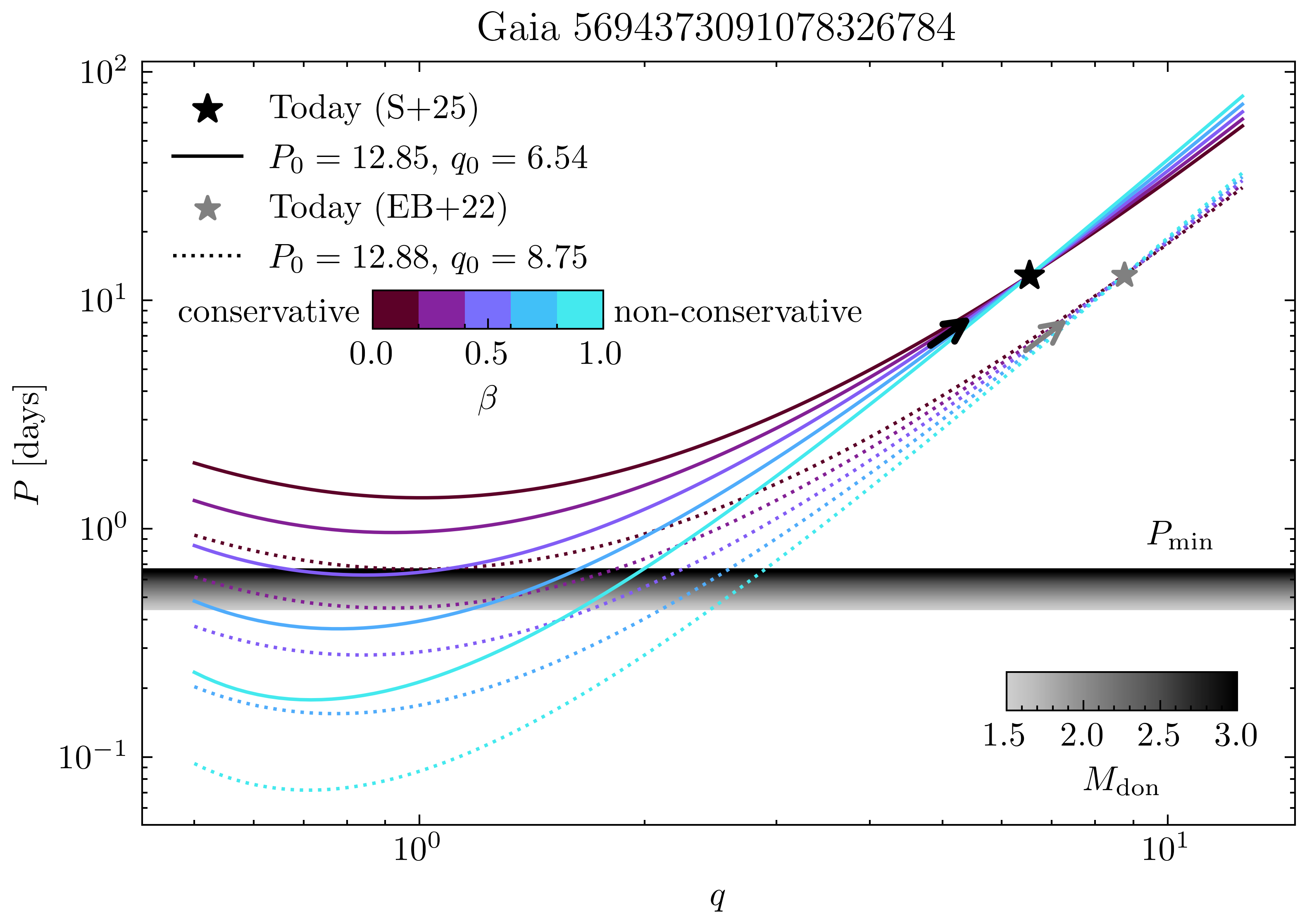}
    \includegraphics[width=\linewidth]{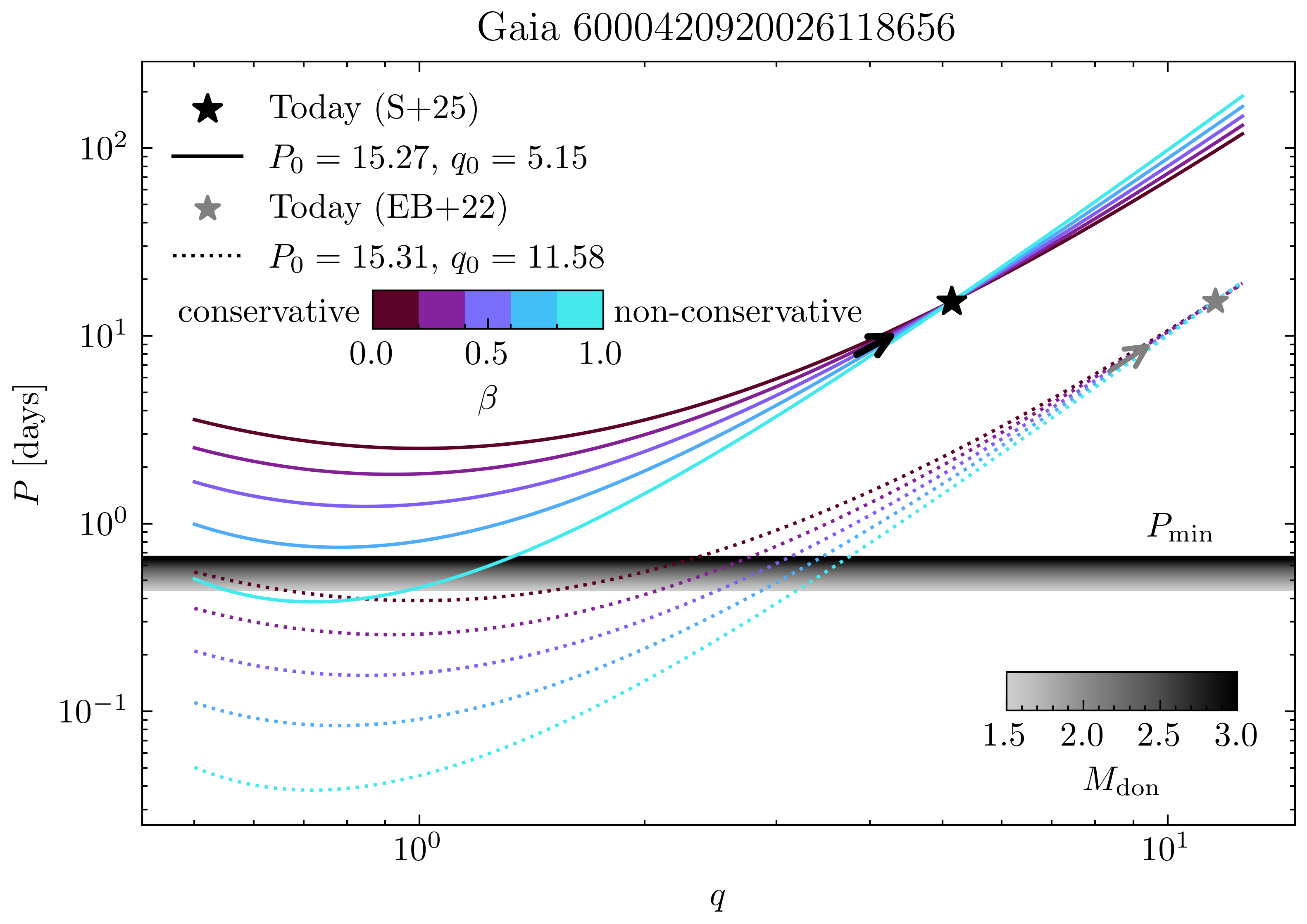}
    
    \caption{Evolution of $P(q)$, subject to the value of $\beta$, for three of the targets, analogous to Figure \ref{fig:masstransfer}.}
    \label{fig:masstransfer_1}
\end{figure}

\begin{figure}[!htb]
    \centering
    \includegraphics[width=\linewidth]{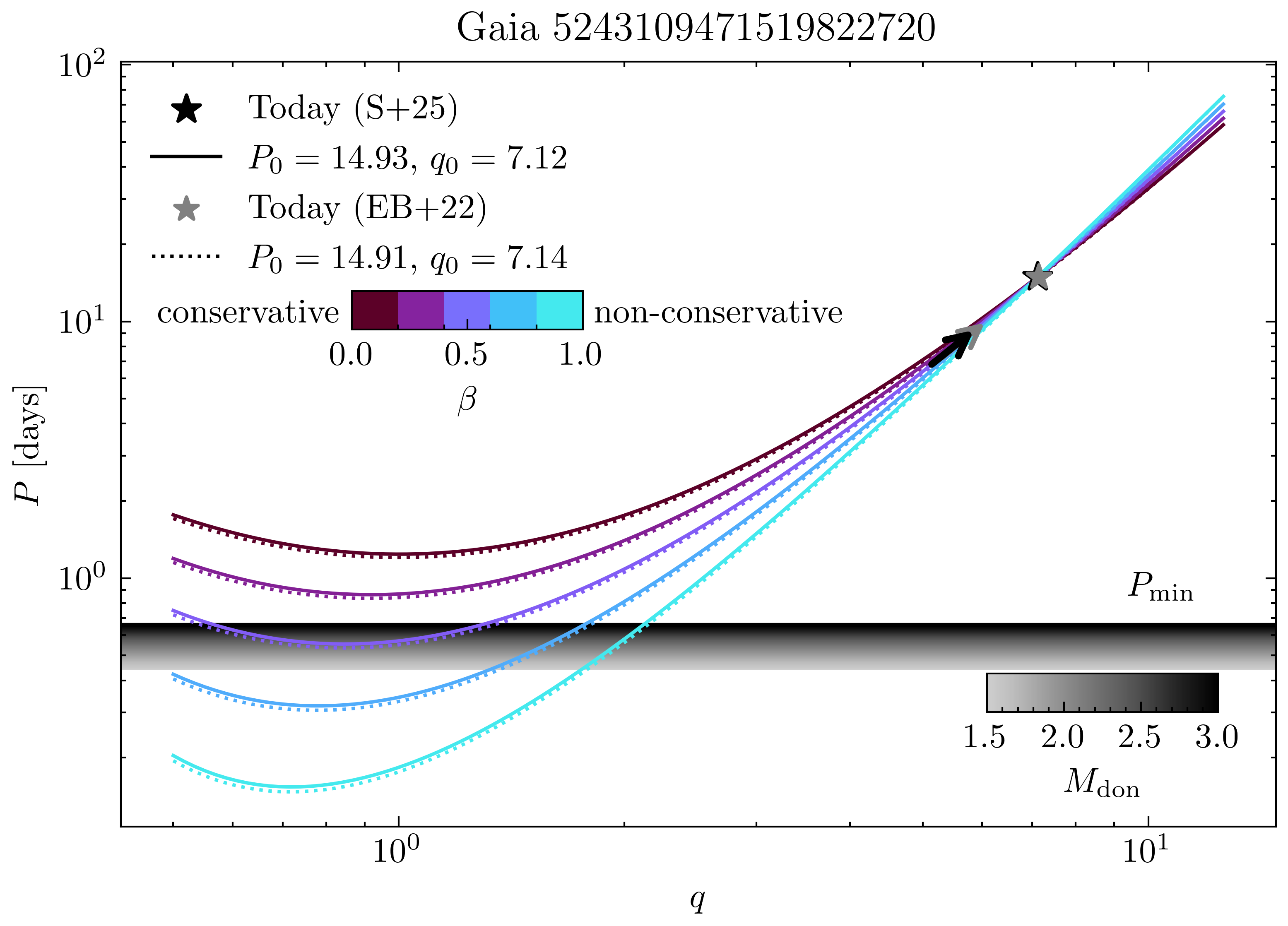}
    \includegraphics[width=\linewidth]{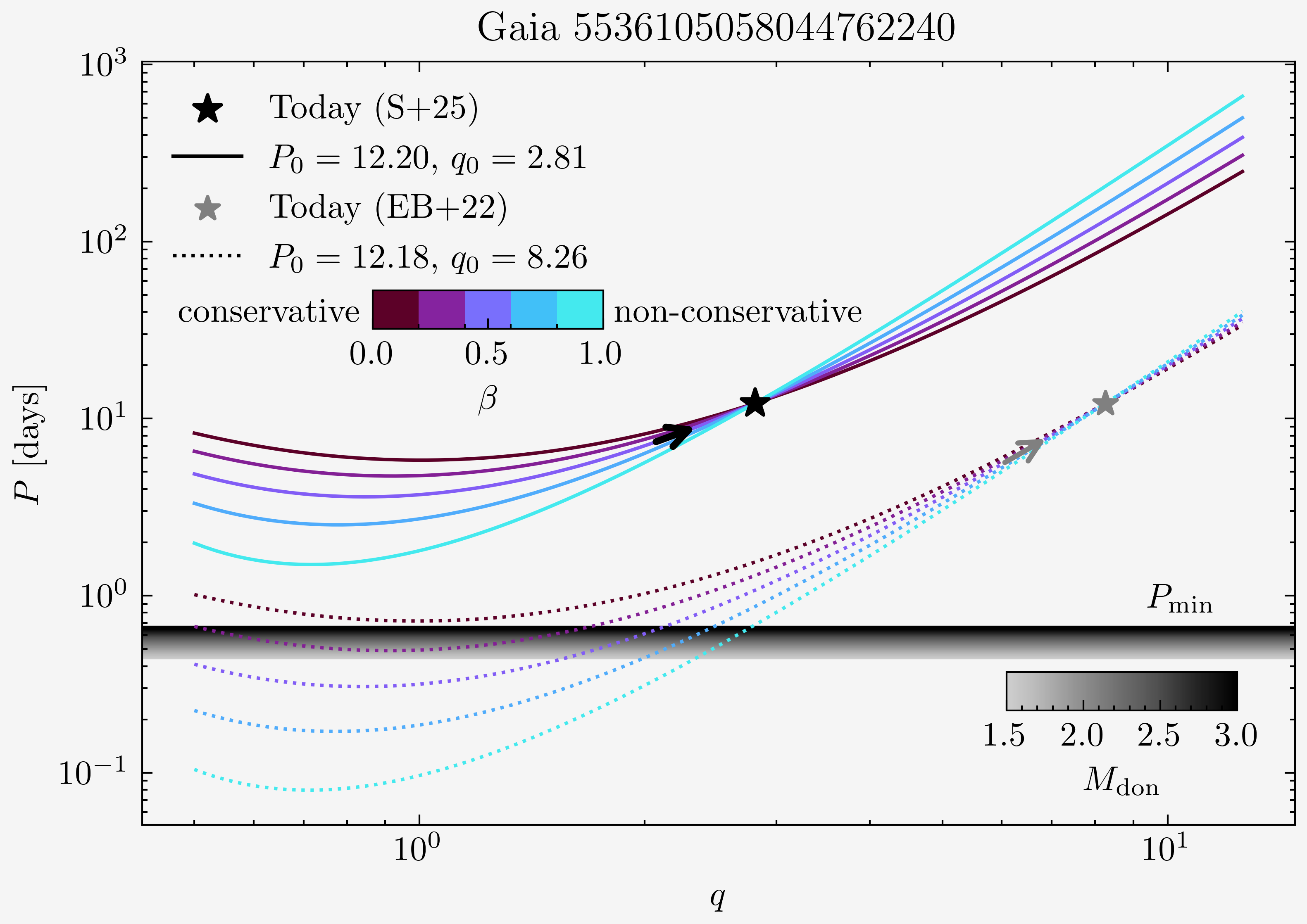}
    
    \caption{Evolution of $P(q)$, subject to the value of $\beta$, for two of the targets, analogous to Figure \ref{fig:masstransfer}. The panel for target G-5536 has a light grey background to highlight the difficulties with its analysis.}
    \label{fig:masstransfer_2}
\end{figure}

Figures \ref{fig:masstransfer_1} and  \ref{fig:masstransfer_2} display the period–mass ratio ($P(q)$) evolution for each binary not shown in the main text, computed using the formalism of \citet{soberman_stability_1997a}.
For each system, we plot the orbital period as a function of mass ratio for different assumptions of the mass-loss parameter $\beta$, assuming $\alpha =  \delta = 0$ (isotropic re-emission).
The grey region marks the minimum orbital period at which the donor would fill its Roche lobe on the main sequence, depending on the assumptions of the initial donor mass.
In our systems, the observed parameters are consistent with relatively conservative mass transfer ($\beta \lesssim 0.7$), in agreement with the conclusions drawn from Section \ref{subsec:MT_history}.

\FloatBarrier
\section{Observation overview}

\begin{table*}[!htb]
\caption{Observational details for the spectra used in this work.}
\begin{tabular}{lllll|lllll}
\hline
Target & MJD & SNR & Exp Time & \verb|ceres| RV & Target & MJD & SNR & Exp Time & \verb|ceres| RV \\ 
 & & & [s] & [km/s] & & & & [s] & [km/s] \\ \hline
G-2933 & 60333.1173 & 103 & 1200 & -14.70 & G-6000 & 60346.3350 & 83 & 1800 & -23.63 \\
G-2933 & 60335.1193 & 89 & 1200 & 60.77 & G-6000 & 60348.3673 & 96 & 1800 & 61.90 \\
G-2933 & 60337.1396 & 103 & 1200 & 111.29 & G-5243 & 60333.2109 & 127 & 1200 & 107.99 \\
G-2933 & 60339.1458 & 100 & 1200 & 102.51 & G-5243 & 60335.2253 & 136 & 1200 & 95.94 \\
G-2933 & 60340.2156 & 102 & 1200 & 73.13 & G-5243 & 60337.2772 & 112 & 1200 & 23.12 \\
G-2933 & 60340.2156 & 102 & 1200 & 73.13 & G-5243 & 60339.2632 & 126 & 1200 & -56.96 \\
G-2933 & 60342.1963 & 100 & 1200 & -4.27 & G-5243 & 60340.2855 & 135 & 1200 & -82.51 \\
G-2933 & 60343.2107 & 83 & 1200 & -39.12 & G-5243 & 60340.2855 & 135 & 1200 & -82.51 \\
G-2933 & 60345.1923 & 83 & 2000 & -68.34 & G-5243 & 60342.2528 & 146 & 1200 & -82.41 \\
G-2933 & 60346.2437 & 65 & 1200 & -58.39 & G-5243 & 60343.3313 & 111 & 1200 & -55.51 \\
G-2933 & 60348.2717 & 97 & 1800 & 1.08 & G-5243 & 60344.2825 & 111 & 1800 & -20.56 \\
G-5694 & 60333.1980 & 67 & 1200 & -22.72 & G-5243 & 60346.3153 & 99 & 1200 & 62.68 \\
G-5694 & 60335.2101 & 73 & 1200 & 76.11 & G-5243 & 60348.3431 & 105 & 1800 & 110.10 \\
G-5694 & 60337.2528 & 76 & 1200 & 154.57 & G-2966 & 60333.0756 & 102 & 1200 & 147.27 \\
G-5694 & 60339.1288 & 71 & 1200 & 147.19 & G-2966 & 60335.0842 & 85 & 1200 & 154.11 \\
G-5694 & 60340.2551 & 76 & 1200 & 103.21 & G-2966 & 60337.0896 & 96 & 1200 & 37.28 \\
G-5694 & 60340.2551 & 76 & 1200 & 103.21 & G-2966 & 60339.0920 & 99 & 1200 & -49.52 \\
G-5694 & 60342.2734 & 73 & 1200 & -3.20 & G-2966 & 60340.1468 & 115 & 1200 & -38.02 \\
G-5694 & 60343.3177 & 64 & 1200 & -42.97 & G-2966 & 60340.1468 & 115 & 1200 & -38.02 \\
G-5694 & 60344.2647 & 61 & 1200 & -57.35 & G-2966 & 60342.1784 & 100 & 1200 & 74.68 \\
G-5694 & 60345.2591 & 73 & 2200 & -47.54 & G-2966 & 60345.1183 & 112 & 2000 & 164.18 \\
G-5694 & 60346.2986 & 55 & 1800 & -13.72 & G-5536 & 60333.1322 & 61 & 1200 & 25.33 \\
G-5694 & 60348.3221 & 52 & 1800 & 86.65 & G-5536 & 60335.1721 & 65 & 1200 & 99.42 \\
G-6000 & 60333.3195 & 123 & 1200 & 71.51 & G-5536 & 60337.1216 & 56 & 1200 & 82.56 \\
G-6000 & 60335.3090 & 96 & 1300 & 110.85 & G-5536 & 60339.1087 & 62 & 1200 & -6.74 \\
G-6000 & 60337.3417 & 77 & 1500 & 82.47 & G-5536 & 60340.2340 & 80 & 1800 & -57.97 \\
G-6000 & 60339.3156 & 91 & 1200 & 2.62 & G-5536 & 60340.2340 & 80 & 1800 & -57.97 \\
G-6000 & 60340.3175 & 112 & 1200 & -41.79 & G-5536 & 60342.2330 & 82 & 1800 & -89.79 \\
G-6000 & 60340.3175 & 112 & 1200 & -41.79 & G-5536 & 60343.2784 & 68 & 1800 & -67.48 \\
G-6000 & 60342.3582 & 136 & 1200 & -100.28 & G-5536 & 60346.2630 & 41 & 1800 & 68.21 \\
G-6000 & 60343.3470 & 108 & 1200 & -104.52 & G-5536 & 60348.2958 & 45 & 1800 & 103.22 \\
G-6000 & 60344.3429 & 74 & 2000 & -91.75 &  & & & & \\
\hline
\end{tabular}
\label{tab:obs}
\end{table*}

Table \ref{tab:obs} summarises the FEROS observations obtained at the MPG/ESO 2.2 m telescope.
For each target and epoch, we list the observing date (MJD), exposure time, and average signal-to-noise ratio per pixel over the observed range, as well as RVs determined by the \verb|ceres| pipeline.
The cadence of typically one to two nights ensured full orbital-phase coverage for systems with periods of approximately ten to twenty days, providing sufficient phase sampling for reliable RV determination and spectral disentangling.

\FloatBarrier
\section{Stellar and system parameters}

\begin{table*}[!htb]
\caption{Stellar parameters for each system from this work}
\centering
\begin{tabular}{llllllll}
\hline
Short ID & $M_\mathrm{ don}$ \tablefootmark{a} & $M_\mathrm{ acc}$ \tablefootmark{a} & $R_\mathrm{ don}$ & $R_\mathrm{ acc}$ & $T_\mathrm{ eff, don}$ & $T_\mathrm{ eff, acc}$ & $d$ \\
 & [M$_{\odot}$] & [M$_{\odot}$] & [R$_{\odot}$] & [R$_{\odot}$] & [kK] & [kK] & [kpc] \\
G-2933 & 0.27 $\pm$ 0.04 & 2.2 $\pm$ 0.2 & 6.02 $\pm$ 0.14 & 2.39 $\pm$ 0.05 & 4.75 $\pm$ 0.25 & 9.50 $\pm$ 0.25 & 1.09 $\pm$ 0.06 \\
G-5694 & 0.24 $\pm$ 0.04 & 2.1 $\pm$ 0.2 & 4.69 $\pm$ 0.07 & 2.58 $\pm$ 0.02 & 4.75 $\pm$ 0.25 & 9.25 $\pm$ 0.25 & 1.65 $\pm$ 0.11 \\
G-6000 & 0.19 $\pm$ 0.03 & 2.2 $\pm$ 0.2 & 5.64 $\pm$ 0.04 & 2.71 $\pm$ 0.03 & 4.75 $\pm$ 0.25 & 9.00 $\pm$ 0.25 & 0.92 $\pm$ 0.07 \\
G-5243 & 0.28 $\pm$ 0.04 & 2.0 $\pm$ 0.2 & 5.46 $\pm$ 0.13 & 2.18 $\pm$ 0.01 & 4.75 $\pm$ 0.25 & 8.75 $\pm$ 0.25 & 0.73 $\pm$ 0.04 \\
G-2966 & 0.23 $\pm$ 0.03 & 2.0 $\pm$ 0.1 & 3.90 $\pm$ 0.02 & 2.22 $\pm$ 0.01 & 5.50 $\pm$ 0.25 & 9.00 $\pm$ 0.25 & 1.16 $\pm$ 0.04 \\
G-5536 & 0.23 $\pm$ 0.04 & 1.9 $\pm$ 0.2 & 4.21 $\pm$ 0.02 & 3.53 $\pm$ 0.02 & 6.75 $\pm$ 0.25 & 7.25 $\pm$ 0.25 & 1.72 $\pm$ 0.04 \\
\hline
\end{tabular}
\tablefoot{\tablefoottext{a}{Stellar masses from \citetalias{el-badry_what_2022}.}}
\label{tab:S_stellar}
\end{table*}

\begin{table*}[h!]
\caption{Stellar parameters for each system from \citetalias{el-badry_what_2022}.}
\centering
\begin{tabular}{llllllll}
\hline
Short ID & $M_\mathrm{ don}$ & $M_\mathrm{ acc}$ & $R_\mathrm{ don}$ & $R_\mathrm{ acc}$ & $T_\mathrm{ eff, don}$ & $T_\mathrm{ eff, acc}$ & $d$ \\
 & [M$_{\odot}$] & [M$_{\odot}$] & [R$_{\odot}$] & [R$_{\odot}$] & [kK] & [kK] & [kpc] \\
\hline
G-2933 & 0.27 $\pm$ 0.04 & 2.2 $\pm$ 0.2 & 7.2 $\pm$ 0.2 & 2.3 $\pm$ 0.2 & 4.4 $\pm$ 0.2 & 9.7 $\pm$ 0.3 & 1.09 $\pm$ 0.02 \\
G-5694 & 0.24 $\pm$ 0.04 & 2.1 $\pm$ 0.2 & 5.5 $\pm$ 0.2 & 1.9 $\pm$ 0.2 & 4.6 $\pm$ 0.2 & 9.9 $\pm$ 0.3 & 1.70 $\pm$ 0.06 \\
G-6000 & 0.19 $\pm$ 0.03 & 2.2 $\pm$ 0.2 & 6.2 $\pm$ 0.2 & 2.5 $\pm$ 0.2 & 4.5 $\pm$ 0.2 & 9.5 $\pm$ 0.3 & 0.92 $\pm$ 0.03 \\
G-5243 & 0.28 $\pm$ 0.04 & 2.0 $\pm$ 0.2 & 6.6 $\pm$ 0.2 & 1.9 $\pm$ 0.2 & 4.6 $\pm$ 0.2 & 9.4 $\pm$ 0.3 & 0.73 $\pm$ 0.01 \\
G-2966 & 0.23 $\pm$ 0.03 & 2.0 $\pm$ 0.1 & 4.7 $\pm$ 0.1 & 2.1 $\pm$ 0.1 & 4.7 $\pm$ 0.1 & 9.4 $\pm$ 0.2 & 1.16 $\pm$ 0.03 \\
G-5536 & 0.23 $\pm$ 0.04 & 1.9 $\pm$ 0.2 & 5.0 $\pm$ 0.2 & 2.5 $\pm$ 0.2 & 4.9 $\pm$ 0.2 & 8.4 $\pm$ 0.3 & 1.77 $\pm$ 0.05 \\
\hline
\end{tabular}
\label{tab:EB_stellar}
\end{table*}

Table \ref{tab:S_stellar} compiles the stellar parameters derived from the disentangled spectra and template fitting described in Section \ref{subsec:stellar_paran}, as well as the radii and distance determined using SED fits as outlined in Section \ref{subsec:sedfit}.
For each system, we report the stellar masses ($M$, taken from \citetalias{el-badry_what_2022}), radii ($R$), effective temperature ($T_{\textrm eff}$), for both the donor and accretor, as well as the distance from the system.

Table \ref{tab:EB_stellar} shows the corresponding values as determined by \citetalias{el-badry_what_2022} and is included for easy comparison of the two works, showing broad consistency of the results.

\begin{table*}[!htb]
\caption{System and orbital parameters for each system as determined in this work.}
\centering
\begin{tabular}{llllllllll}
\hline
Short ID & Gaia $G$mag \tablefootmark{a}& Gaia parallax \tablefootmark{a}& $P_\mathrm{ orb}$ \tablefootmark{b} & K1 & $P_\mathrm{ orb}$ & $M_0$ & $e$ & $\omega$ & v$_\mathrm{ COM}$ \\
 & [mag] & [mas] & [days] & [km/s] & [days] & [rad] & & [rad] & [km/s] \\
 \hline
G-2933 & 11.07 & 0.92 & 14.71 & 92.1 & 14.70 & 2.75 & 0.01 & 1.76 & 23.9 \\
G-5694 & 11.91 & 0.59 & 12.88 & 109.6 & 12.85 & 2.54 & 0.02 & 2.39 & 52.6 \\
G-6000 & 10.73 & 1.08 & 15.31 & 107.6 & 15.27 & 3.27 & 0.01 & 3.25 & 3.7 \\
G-5243 & 10.45 & 1.37 & 14.91 & 101.4 & 14.93 & 3.77 & 0.01 & 2.34 & 10.0 \\
G-2966 & 11.20 & 0.86 & 10.40 & 110.4 & 10.43 & 3.64 & 0.02 & 2.09 & 59.5 \\
G-5536 & 12.15 & 0.57 & 12.18 & 97.9 & 12.20 & 3.00 & 0.02 & 3.28 & 6.7 \\
\hline
\end{tabular}
\tablefoot{\tablefoottext{a}{G-band magnitudes and Gaia DR3 parallaxes from \citet{gaiacollaboration_gaia_2023c}.} \tablefoottext{b}{Comparative orbital periods from \citetalias{el-badry_what_2022}.}}
\label{tab:orbital}
\end{table*}

Table \ref{tab:orbital} compiles the orbital and stellar parameters derived in this work.
For each system, we provide the velocity semi-amplitude of the primary, $K1$; orbital period, $P$; mean anomaly, $M_0$; eccentricity, $e$; argument of pericentre, $\omega$; and systemic velocity, $v_\mathrm{ COM}$; obtained from the nested-sampling orbital fits described in Section \ref{subsec:orbit}.
These periods combined with the mass ratios obtained using TODCOR in Section \ref{subsec:TODCOR} form the quantitative basis for the discussion of the system's mass transfer history presented in Section \ref{subsec:MT_history} and \ref{sec:discussion}.

\end{appendix}

\end{document}